\documentclass[fleqn,usenatbib]{mnras}
\usepackage[T1]{fontenc}

\DeclareRobustCommand{\VAN}[3]{#2}
\let\VANthebibliography\thebibliography
\def\thebibliography{\DeclareRobustCommand{\VAN}[3]{##3}\VANthebibliography}

\usepackage{graphicx}	
\usepackage{amsmath}	
\usepackage{amssymb}
\usepackage[flushleft]{threeparttable}
\usepackage{hyperref}
\usepackage{subfigure}

\newcommand{\kms}{\mbox{km s$^{-1}$}}
\newcommand{\vlos}{V_{\rm los}}

\newcommand{\unsim}{\mathrm{\sim}}
\newcommand{\mtol}{$M_{*}/L_{\rm v}$}
\newcommand{\mtolunit}{\rm M_\odot/\rm L_\odot}
\newcommand{\veldis}{\rm \sigma_{*}}
\newcommand{\ppxf}{\texttt{PPXF}}
\newcommand{\minus}{\scalebox{0.75}[1.0]{$-$}}
\def\degr{\hbox{$^\circ$}}
\def\arcmin{\hbox{$^\prime$}}
\def\arcsec{\hbox{$^{\prime\prime}$}}

\def\HII{{{\sc H\thinspace ii}}}
\def\Ha{H$\alpha$}
\def\Hb{H$\beta$}

\title[Stars and ionized gas in UGCA~320]{Stars and ionized gas in UGCA~320: a nearby gas-rich, dwarf Irregular galaxy}

\author[Alabi et al.]{
Adebusola B. Alabi,$^{1}$\thanks{E-mail: abbgstar@gmail.com }
S. Ilani Loubser,$^{1,2}$
Moses K. Mogotsi,$^{3,4}$
N. Zabel$^{5}$
\\
$^{1}$Centre for Space Research, North-West University, Potchefstroom Campus, 2520, South Africa\\
$^{2}$National Institute for Theoretical and Computational Sciences (NITheCS), Potchefstroom 2520, South Africa\\
$^{3}$South African Astronomical Observatory, P.O. Box 9, Observatory, 7935, Cape Town, South Africa\\
$^{4}$Southern African Large Telescope Foundation, P.O. Box 9, Observatory, 7935, Cape Town, South Africa\\
$^{5}$Department of Astronomy, University of Cape Town, Private Bag X3, Rondebosch 7701, South Africa\\
}

\date{Accepted XXX. Received YYY; in original form ZZZ}

\pubyear{\the\year{}}

\begin{document}
\label{firstpage}
\pagerange{\pageref{firstpage}--\pageref{lastpage}}
\maketitle

\begin{abstract}
UGCA~320 is a gas-rich dwarf irregular galaxy which belongs to a nearby, relatively isolated group of dwarf galaxies. Here, we combine multi-band \textit{HST} imaging data with deep long-slit SALT/RSS and integral-field VLT/MUSE spectral data to study the stellar and ionized gas components of UGCA~320. Our imaging data analysis reveals a very blue ($V-I\unsim0.1$~mag), flattened radial colour profile. We detect an abundance of ionized gas in UGCA~320 powered mostly by recent star formation. The stellar disc in UGCA~320 is populated predominantly by young ($\unsim120$~Myr) and metal-poor ($\unsim15-30$ per cent solar metallicity) stars and it rotates in the same sense as the ionized gas disc but with higher rotation velocities, and possibly in different planes. Our analysis reveals a sharp transition in the kinematic properties of the discs at radius $\unsim10\arcsec$ 
($\unsim0.3$~kpc) and distortions in the outer disc region. We show that these features are consistent with a recent tidal interaction most likely with its close neighbour -- UGCA~319. We discuss our results in the context of interacting dwarf galaxies and also show that similar inferences can be made independently from the long-slit data analysis as with the integral-field data.  
\end{abstract}


\section{Introduction}
Dwarf galaxies, being the least luminous, most abundant and diverse galaxy type, play an important role in our understanding of the Universe. For example, they have recently been reported to be responsible for most of the flux that reionized the early Universe \citep{Atek2024}. They are believed to be the building blocks of more massive present-day galaxies within the hierarchical structure formation paradigm of the $\Lambda$CDM cosmology, where galaxies assemble over time experiencing processes such as galaxy interactions and mergers. In the Local Group, the Large Magellanic Cloud (LMC) and the Small Magellanic Cloud (SMC) are perhaps the best-known example of gas-rich, dwarf Irregular (dIrr) galaxy pair undergoing dwarf-dwarf galaxy interaction \citep{Besla2012, Diaz2012}. The presence of the Magellanic Stream, the Magellanic Bridge, the leading arm and the peculiar shape of the stellar kinematics are incontrovertible signatures of the recent interactions between these prototype Magellanic Irregulars. Similar tidal features are expected in interacting systems of $\le L_{\odot}$ galaxies \citep{Toomre1972, Barnes1988}. Gas-rich dwarf galaxy interactions and mergers are known to disrupt gas rotation in discs and trigger bursts of star formation, which may in turn heat up the interstellar medium thus preventing gas cooling, alongside the supernovae feedback which may lead to gas outflows \citep{Mihos1996, Matteo2005}. \citet{Stierwalt2015}, for example, showed that there is a factor of two enhancement in the star formation rate of interacting dwarf galaxy pairs that are closer than $50$~kpc.

We have identified a dwarf galaxy -- UGCA~320 -- from the ongoing MHONGOOSE\footnote{MeerKAT HI Observations of Nearby Galactic Objects: Observing Southern Emitters} survey \citep{deBlok2024} as a prime candidate where these processes and their complex interplays can be studied. UGCA~320, also known as DDO~161 or HIPASS~J1303-17b, is a gas-rich, optically blue, dwarf irregular galaxy with two spectroscopically confirmed dwarf neighbours -- UGCA~319 and LEDA~886203. \citet{Karachentsev2017} originally identified UGCA~320 and UGCA~319 as an isolated galaxy pair with a physical projected separation of $32.7$~kpc in the plane of the sky and a line-of-sight velocity ($\vlos$) separation of $\unsim16\ \kms$. \citet{Healy2024} recently reported that LEDA~886203, at an angular separation of $\unsim27\arcmin$ from UGCA~320\footnote{This corresponds to a physical projected separation of $\unsim47$~kpc assuming it is at the same distance as UGCA~320. Unlike UGCA~320 and UGCA~319, with well-constrained distance measurements in the literature, the precise distance of LEDA~886023 is unconstrained.}  has a $\vlos$ measurement of $729$~\kms\ (from their deep HI study), implying a velocity separation of $\unsim10\ \kms$. UGCA~320 is the most prominent and gas-rich galaxy in this system of three dwarf galaxies. Furthermore, based on their proximity in velocity-position phase plane, it is plausible to assume that they might be interacting -- one of the themes we explore in-depth later in this work. The baryonic budget in UGCA~320 is dominated by the HI component, i.e., M$_{\rm HI}$/M$_{*}{\boldsymbol{ \unsim 10}}$. Also, the HI extent is at least $2\times$ larger than the optical stellar disc \citep{Cote2009}. Yet despite its huge gas reservoir, the published star formation rate in the literature for UGCA~320 is at best modest for its gas mass, e.g., $\le0.02~M_{\odot}$ yr$^{-1}$  \citep{Meurer2006, Cote2009, Leroy2019}.   

\begin{figure*}
\includegraphics[width=0.99\textwidth]{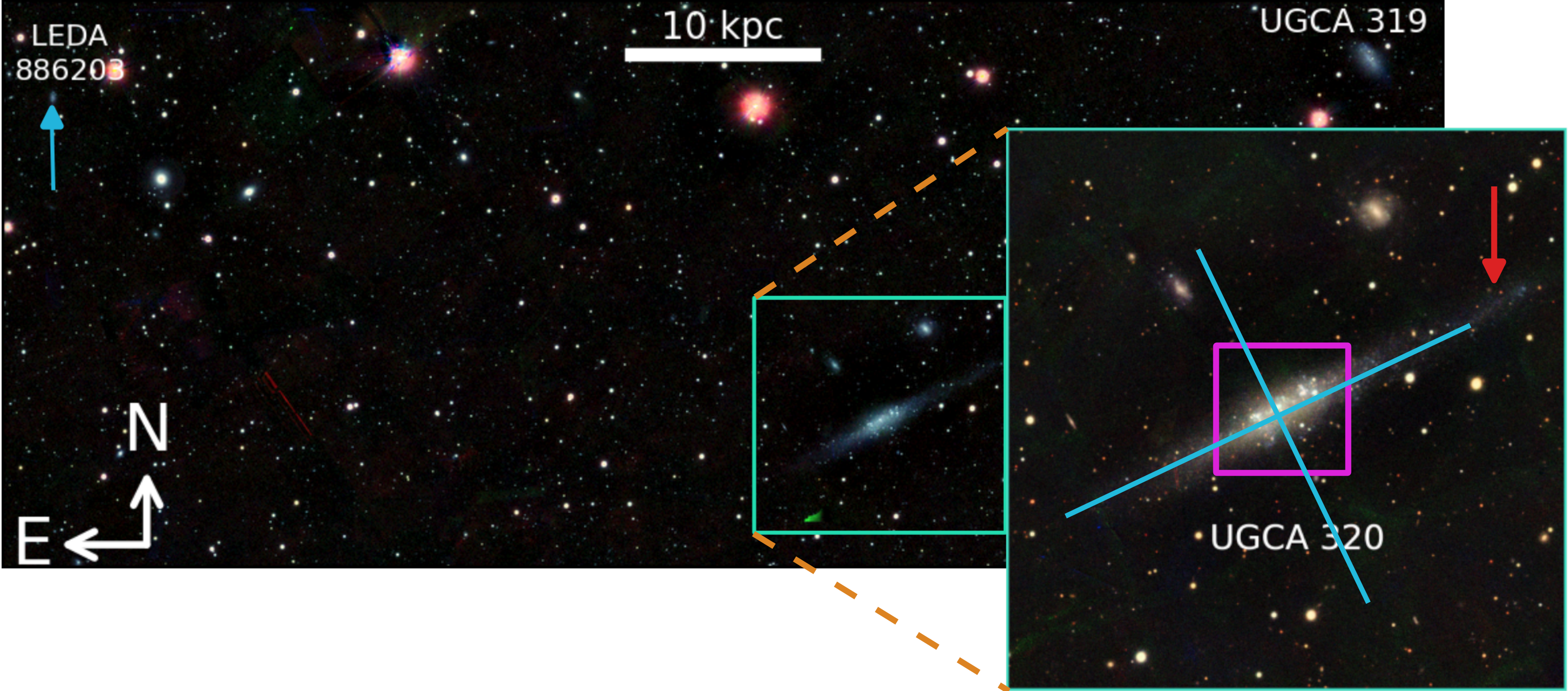}
\caption{A Pan-STARRS $grizy$ composite image of the field of UGCA~320. The background image is $78$~kpc $\times$ $31$~kpc, North is up and East is left. UGCA~320 is the prominent, elongated galaxy in the background image, highlighted in the green box. The two neighbours, LEDA~886203 and UGCA~319 are labelled and can be seen in the top-left (top of the blue arrow) and top-right portions of the background image, respectively. The foreground image (bigger green box) is a zoom-in view of the immediate neighbourhood of UGCA~320 with the VLT/MUSE field-of-view (purple box) and SALT long-slit positions (skyblue lines) overlaid. A faint, asymmetric stellar extension is visible along the major axes of UGCA~320, in the north-west direction, marked with the red arrow.}
\label{fig:colImg}
\end{figure*}

In this work, we use modern optical spectroscopic data to re-measure the star formation rate in UGCA~320 and re-examine the reason(s) for its low star formation efficiency in spite of its gas-richness. The stellar population parameters of UGCA~320, i.e., age and metallicity, have not been investigated in the literature, as is the case for most low-surface brightness galaxies due to their faintness. \citet{Lee2003}, however, obtained a sub-solar ionized gas-phase metallicity of $\mathrm{12 + log (O/H)} \unsim 8.06$ for UGCA~320 from their long-slit observations. We will perform a joint kinematics and chemical enrichment analysis of the stars and ionized gas components of UGCA~320, using state-of-the-art full spectrum fitting technique and spectral libraries, and deep spectral data from $\unsim10$~m class telescopes. We summarize the properties of UGCA~320 in Table~\ref{tab:globprop} and show a composite image of the field of UGCA~320 in Fig.~\ref{fig:colImg}.

\begin{table}
\caption{Global Parameters for UGCA~320}
\label{tab:globprop}
\begin{threeparttable}
\footnotesize
\begin{tabular}{lcc} \hline
Parameter & Value & Ref \\
\hline
Other names       &      DDO~161, PGC045084,  &       1 \\
                  &      HIPASS~J1303-17b     & 		   \\
D (Mpc)           &      $6.03\pm0.25$                           &       2 \\
RA, Dec. (J2000)  &      13:03:16.74, -17:25:22.9               &       1 \\
$B_{T}$, (mag)    &      $13.5$                                  &       3 \\  
Position Angle    &      $116.9\degr^{a}$                        &       3 \\  
$b/a$             &      $0.17$                                  &       3 \\
Inclination       &      $83\degr$                               &       4 \\
log(M$_{*}$/M$_{\odot}$) &      $7.91^{b}$                       &       4 \\  
log(M$_{\rm HI}$/M$_{\odot}$)&      $8.97^{b}$                      &       4 \\
$V_{\rm sys} (\kms)$    &      $739.9$                           &       4 \\ \hline
\end{tabular}
\begin{tablenotes}
\small
\item (1) NASA Extragalactic Database; (2) \citet{Karachentsev2017}; (3) HyperLEDA \citet{Makarov2014}; (4) \citet{deBlok2024}
\item $a$ HyperLEDA quotes an average position angle of $109\degr$, however, we note that there is an outlier in their compilation responsible for their lowered quoted value. 
\item $b$ While we have adopted the stellar mass from \citet{deBlok2024} in this table, in the analysis that follows, we note that published stellar mass for UGCA~320 spans a wide range, up to log(M$_{*}$/M$_{\odot})\unsim8.75$, in the literature.
\end{tablenotes}
\end{threeparttable}
\end{table}

This work is organised thus. Section~\ref{obs_data} describes the imaging and spectroscopic data used in this work. Section~\ref{data_analysis} describes our analysis of the spectral data in detail. In Section~\ref{results} we present the results of our analyses. We discuss the implications of our results and summarize the key findings and outstanding questions in Section~\ref{discuss}. Throughout this work, we adopt an $H_{0} = 73 \pm 5~\kms \rm{Mpc}^{-1}$ and at the distance of UGCA~320, $1\arcsec$ corresponds to $\unsim30$~pc.

\section{Observations and data reduction}
\label{obs_data}
\subsection{Archival imaging data }
\label{sec_HST}
\citet{Karachentsev2017} published broad-band $HST$/ACS imaging data of UGCA~320 in the F606W ($V$) and F814W 
($I$) filters. The authors used the tip of the red giant branch method to determine that the galaxy is at a distance of $6.03$~Mpc. We retrieved these data from the Mikulski Archive for Space Telescopes (MAST) Portal\footnote{\url{https://mast.stsci.edu}} of the Space Telescope Science Institute (STScI). Visual inspection of these $HST$ images suggests that the starlight in UGCA~320 extends asymmetrically beyond the limits of the ACS field-of-view (FoV). We confirm this by inspecting archival wide-field $R$-band imaging of UGCA~320 available in the literature, e.g., SINGG collaboration: \citet{Meurer2006}; \citet{Cote2009}. UGCA~320 has a faint stellar region that extends for $\unsim100 \arcsec$ in the north-west direction along the major axis with six \HII\ regions identified by \citet{Cote2009} - see their fig.~1. We obtain $V-I$ radial colour profile from this $HST$ data and present the result in Section~\ref{radial_col}.

\subsection{SALT RSS Long-slit observations}
UGCA~320 was observed with the Robert Stobie Spectrograph (RSS) on the Southern African Large Telescope (SALT) in Sutherland, South Africa over five nights between February - June, 2021 as part of the observing programs PI: Mogotsi; 2020-2-SCI-029 and 2021-1-MLT-002. The observations were performed in the long-slit mode with two setups, i.e., the low resolution (hereafter LR) setup observed with
the PG0900 grating and the high resolution (hereafter HR) setup observed with the PG2300 grating. The two setups yield spectra ranging from $4060 - 7120$ \AA\ and $5880 - 6730$ \AA\ with full width at half maximum (FWHM) spectral resolution of $4.8~\&~1.3$ \AA\ at the corresponding central wavelengths, respectively. The LR and HR SALT longslits have sizes $1.25\arcsec \times 8\arcmin$ and $0.6\arcsec \times 8\arcmin$, respectively. 

We observe the target galaxy along two slit position angles, $116.9\degr$ and $26.9\degr$, corresponding to the major and minor axes orientation 
of the imaging data, respectively. The science data were obtained in fair seeing conditions $\le 2\arcsec$ with exposure times ranging from $2\times1000$~s to $2\times1920$~s. We note that absolute flux calibration is not possible with SALT spectroscopic data due to the limitation imposed by the strongly varying pupil size of the telescope\footnote{\url{https://pysalt.salt.ac.za/proposal\_calls/current/ProposalCall.pdf}}. However, for each observation, we perform relative flux calibration using spectra from standard stars taken with the same setup.  

\begin{figure}
\includegraphics[width=0.48\textwidth]{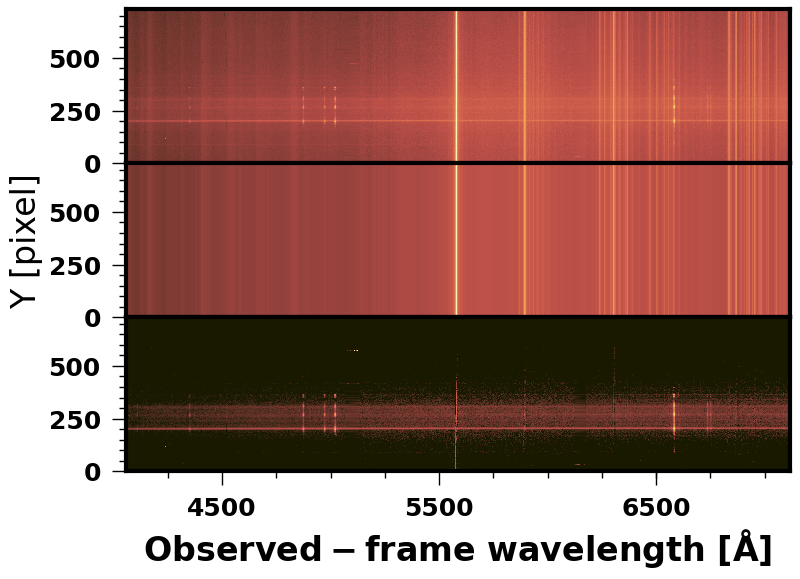}
\caption{\textbf{Top panel: }Two-dimensional galaxy spectrum of UGCA~320 obtained along the major axis in a $\unsim1000$~s exposure \textit{before} sky subtraction. \textbf{Middle panel: }2D-Sky model constructed using the optimal sky subtraction routine presented in \citet{Kelson2003}. \textbf{Bottom panel: }2D Sky-subtracted galaxy spectrum used in the analysis presented in this work. Residuals can still be seen around the strong $\unsim5570$ \AA\ sky feature which we mask out in our analysis.}
\label{fig:img2d}
\end{figure}

Our SALT RSS long-slit data was reduced with the standard SALT science pipeline. We then determined and applied a wavelength solution using arc lamp spectra observed immediately after all science frames. We used the L.A. Cosmic algorithm \citep{vDokkum2001} to remove Cosmic rays from the two-dimensional (2D) spectral data. We performed sky subtraction using the optimal sky subtraction routine (which we describe briefly below) presented in 
\citet{Kelson2003} and median combined the individual frames to obtain our final 2D spectra (an example of which is shown in Fig.~\ref{fig:img2d}) for each long-slit configuration. 

Our implementation of the sky subtraction routine includes oversampling pixels in the sky-defined rows of the 2D galaxy plus sky spectral frame by a factor of $20$. The oversampled 2D sky subgrid is then fit with low-order polynomials to obtain a subgrid sky model which we re-project onto the entire 2D spectral grid. We subtract this re-projected 2D sky model to obtain our sky-subtracted 2D spectral data for each science observation, which are then median combined. We show in Fig.~\ref{fig:img2d} an example 2D galaxy plus sky spectral frame, the 2D sky model and the sky-subtracted 2D frame. Finally, we correct for Galactic extinction using the extinction maps of \citet{Schlafly2011} and the extinction law of \citet{Cardelli1989} with $R_{V} = 3.1$.

\subsection{VLT/MUSE IFU spectroscopy}
An archival, science-ready MUSE data cube covering UGCA~320 is publicly available as part of the observing program 
PI: Bian; 105.20GY. UGCA~320 was observed on April 22, 2021, with a total exposure time of $4480$~s. The data was acquired in the 
Wide Field Mode with a FoV of $1 \arcmin \times 1 \arcmin$, a spatial resolution of $0.2 \arcsec$ per pixel, 
with a spectral range spanning $4750 - 9350$ \AA\ in steps of $1.25$ \AA\ and a median FWHM spectral resolution of $\unsim2.6$ \AA. 
The average seeing during the observations is $0.75 \arcsec$ which implies a spatial resolution of $\unsim22$~pc at the 
distance of UGCA~320. Likewise, each pixel covers $\unsim34$~pc$^{2}$ and the FoV covers $\unsim1.7 \times 1.7$~kpc. 
We retrieved the wavelength and flux calibrated, sky-subtracted phase 3 data cube from the ESO Archive Science 
Portal\footnote{\url{https://archive.eso.org/scienceportal/home}} that was reduced using the standard reduction 
pipeline \citep{Weilbacher2020}. We correct the data cube for Galactic extinction in the same way as the SALT 2D spectral data.

\section{Data Analysis}
\label{data_analysis}
\subsection{Extracting 1D spectra from 2D SALT long-slit spectra}
\label{gal_centre}
\begin{figure*}
\includegraphics[width=0.98\textwidth]{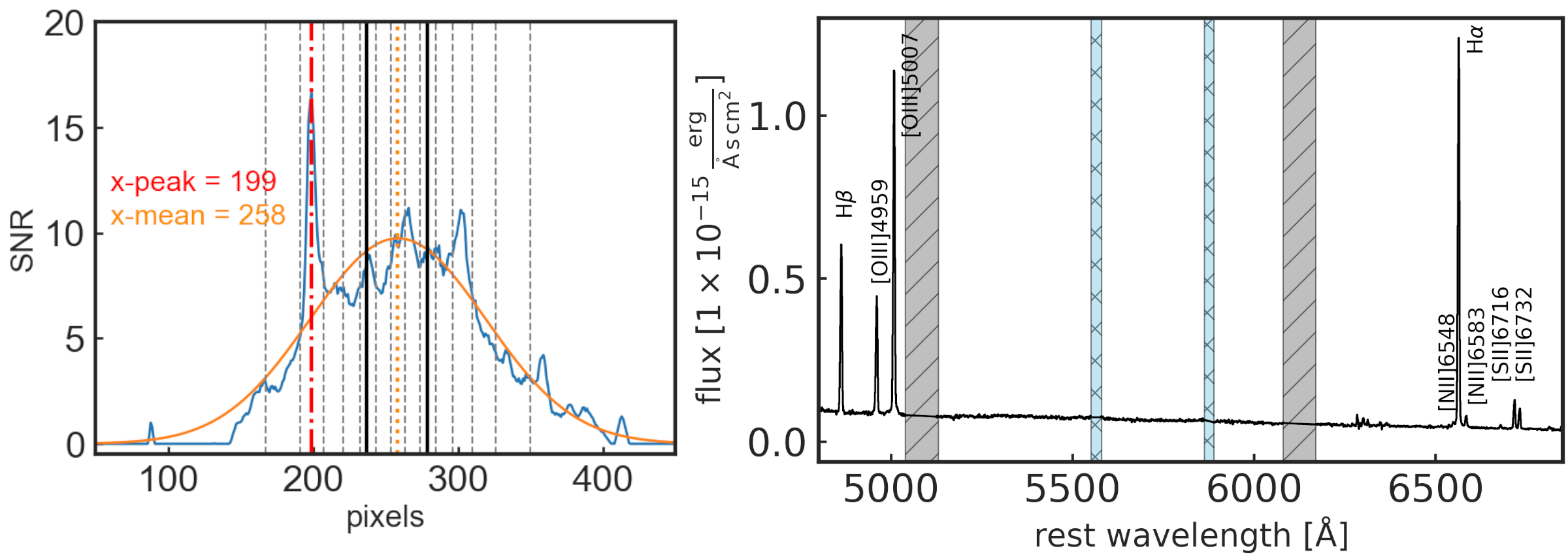}
\caption{\textbf{Left panel: } Collapsed sky-subtracted 2D spectrum showing the light profile (blue colour) from which we extract a central 1D spectrum (within the black solid lines) and binned spectra from adaptively determined pixel rows (grey dashed lines). The orange curve is a Gaussian fit to the light profile, centred on the orange dashed line. Note that the brightest peak along the slit (dashed red line) does not correspond to the centre of the light profile. \textbf{Right panel: } Example 1D spectrum extracted from the central region showing strong emission features. The grey and skyblue shaded regions correspond to the SALT CCD chip gaps and residual sky regions, respectively, which we mask out from our spectral analyses.}
\label{LS_spec}
\end{figure*}

From each 2D sky-subtracted spectrum, we make two types of spectral extractions after collapsing the 2D spectrum along the wavelength axis: first, a single global one-dimensional 
spectrum where we maximise the signal-to-noise ratio (SNR), and second, multiple spectra from an adaptively determined number of pixel rows to make radial profiles.
The first extraction is typically from the central $30$ pixel rows which correspond to $\unsim8 \arcsec$ ($\unsim0.3$~kpc).
The second extraction is optimized to reach the SNR required for our stellar analysis.
Due to the clumpy nature of the \HII\ knots prevalent along the long-slit, the brightest pixel does not necessarily coincide with the peak of the 
Gaussian profile or the galaxy centre. We therefore fit a Gaussian profile to the collapsed light profile, masking out the sharp 
peaks from \HII\ emission regions and binning the profile so that the area under each segment is the same. This ensures that the 
extracted spectra, even in the outskirt regions, have sufficient SNR $\ge10$ per \AA\ in the stellar continuum required for our subsequent analyses. 
The number of pixel rows summed over to get our binned 1D spectra increases from the centre of the spatial profile to the outskirts. In the 
brightest central region of the collapsed light profile, we typically make spectral extractions in bins spanning $8$~pixels ($\unsim2 \arcsec$), 
while in the outskirts, we sum over $22$~pixels ($\unsim6 \arcsec$) out to a projected radii of $\unsim20 \arcsec$ and $\unsim10 \arcsec$ 
along the major and minor axis from the galaxy centre, respectively. In the case of the data from our HR setup, we 
are only able to make the first single extraction, and only from the major axis observation, due to SNR constraints. Fig.~\ref{LS_spec} shows 
the collapsed light profile from an example slit and its corresponding 1D global spectrum. 

\subsection{Spectral fitting of SALT long-slit spectra}
We employed the commonly used penalized pixel fitting package (\ppxf; \citealt{Cappellari2004,Cappellari2017}) 
to extract the stellar kinematics (line-of-sight velocity, $\vlos$ and velocity dispersion, $\veldis$), stellar population parameters 
(age, metallicity, stellar mass-to-light ratio ($M_{*}/L_{\rm v}$)), and emission-line parameters (flux and kinematics) from our SALT long-slit 1D spectra. Throughout this work, we only measured $\veldis$ from the HR setup, and only from the centrally extracted spectrum along the major axis due to SNR limitations. The HR setup has a FWHM spectral resolution of $1.3$ \AA\ at $6300$ \AA\ corresponding to an instrumental dispersion of $\unsim 26~\kms$, adequate to measure the $\veldis$ of UGCA~320. On the other hand, the LR setup (FWHM of $4.8$ \AA\ at $5925$ \AA) and MUSE data (FWHM of $2.6$ \AA\ at $5925$ \AA) have instrumental dispersions of $\unsim110~\& \unsim60~\kms$, respectively. These spectral resolutions are too low compared to the expected $\veldis$ of UGCA~320, given its stellar mass. We shifted the spectra to rest-frame wavelengths using the systemic velocity from Table~\ref{tab:globprop} and logarithmically rebinned them to conserve flux. We fitted the full spectral data within the $4850-7000$ \AA\ wavelength range but masked regions with strong sky residual features, i.e., $\unsim5570$ \& $5587$ \AA, and the two chip-gap regions. 

As recommended by \citet{Cappellari2017}, we performed the full spectrum fitting in two steps: (i) stellar kinematics and stellar 
population parameters analyses, where we additionally mask out gas emission-line regions, and (ii) emission-line parameters analysis, where we fix the stellar kinematics to the results from the previous step and simultaneously fit the stellar continuum and the emission line features. Our HR spectral data, despite being well-suited for studying the stellar kinematics of dwarf galaxies due to its high spectral resolution, is limited in its utility for determining stellar population properties due to a lack of age and metallicity sensitive features in the covered wavelength range. We therefore exclude it from our stellar population parameter analyses.
  
We used template spectra from the Flexible Stellar Population Synthesis (FSPS; \citealt{Conroy2009, Conroy2010}) and the GALAXEV (\citealt{Bruzual2003}) models to fit our LR spectral data independently. We have chosen to use these two SPS models, particularly for our stellar population analysis, since the models are based on different stellar components, often yielding discrepant stellar ages and metallicities. The former of these two SPS models is based on the MILES stellar library (\citealt{Vazdekis2015}) while the latter is based on the STELIB spectral library (\citealt{leBorgne2003}). Both models are computed with the Salpeter initial mass function but use different isochrones, i.e., FSPS models are generated with the MIST isochrones (\citealt{Choi2016, Paxton2019}) and GALAXEV with the Padova Isochrones (\citealt{Bressan1993, Girardi2000}). 
The two models cover a large and similar range of ages ($0.001 - 15.8$~Gyr) and metallicities ($\rm [Z/H]$ from $-1.75$ to $0.25$~dex) and have the same FWHM spectral resolution of $\unsim2.5$ \AA\ and spectral sampling of $0.9$ \AA\ per pixel over the wavelength range of our LR data. We employed template spectra from the P{\'E}GASE-HR model to fit our HR long-slit data. They are based on high-resolution (FWHM spectral resolution of $0.55$ \AA\ at $5500$ \AA) stellar spectra from the ELODIE spectral library (\citealt{Prugniel2001, Prugniel2004}) and the Padova isochrones. We convolved the template spectra to the spectral resolution of our SALT data before performing the fit.

For the emission-line analysis, each fitted emission feature is represented by a single Gaussian profile within the \ppxf~framework after isolating the stellar continuum. The fitted features include the strong recombination Balmer lines (\Hb, \Ha) and the forbidden lines ([OI]$\lambda6300,6364$, [OIII]$\lambda\lambda4959,5007$ (hereafter [OIII]), [NII]$\lambda\lambda6548,6583$ (hereafter [NII]) \& [SII]$\lambda\lambda6717,6732$ (hereafter [SII]\footnote{[SII] is evaluated as the sum of the flux from the doublet.}). Ionized gas flux amplitudes are allowed to vary freely but are subject to constraints on flux ratios from atomic physics \citep{Osterbrock1989}. Ionized gas kinematics, however, is constrained such that all recombination lines share the same kinematics and all the forbidden lines also share the same kinematics, following the method described in \citet{Sarzi2006}. The flux measurements are dereddened to remove the effects of internal dust extinction using the expected intrinsic ratio of \Ha~/\Hb$=2.86$, i.e., the Balmer decrement, assuming case B recombination with a temperature $T=10^{4}$~K and an electron density $n_{e} = 10^{2}\ \rm cm^{-3}$~(\citealt{Osterbrock1989}), and the \citet{Cardelli1989} reddening law where $R_{V} = 3.1$\footnote{We used the \texttt{Python}-based extinction package from \url{https://github.com/kbarbary/extinction}}.

\subsection{Spectral fitting of MUSE spectra}
\label{spec_fittx}
Here we detail how we analysed the MUSE data cube to obtain spatially resolved stellar kinematics (only $\vlos$), stellar population parameters, 
and emission-line properties. While our spectral fitting method is the same as with the SALT long-slit data, i.e., we use the 
\textsc{ppxf} full-spectrum fitting code and the FSPS and GALAXEV models to fit the data in rest-frame separately, we highlight the 
differences in implementation below. We spatially binned the data cube to increase the SNR before fitting for the stellar parameters 
of interest. We used the Voronoi binning method (\citealt{Cappellari2003}) to achieve an average SNR of $25$ per \AA\ in the continuum over the wavelength range $4850 - 7000$ \AA, excluding individual spaxels with SNR below a threshold of $3$ per \AA. 
We note that we perform full spectral fitting for both MUSE and SALT data over the same wavelength range since the method is well-known 
to be sensitive to the analyzed spectral range (\citealt{Koleva2008}). For the emission-line analysis, we perform spectra fitting on 
individual spaxels to maximise the spatial sampling of the emission-line maps, again excluding spaxels below our SNR threshold. The template spectra used in our MUSE full spectrum fitting were all convolved to the wavelength-dependent spectral resolution of MUSE obtained from the relation in \citet{Bacon2017}. 

\begin{figure}
\includegraphics[width=0.48\textwidth]{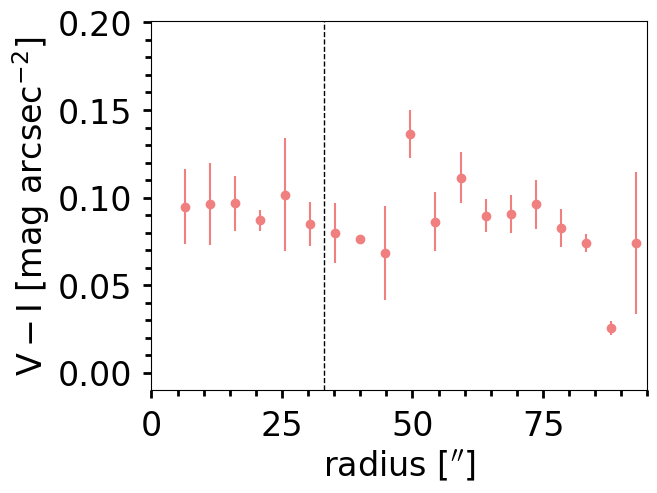}
\caption{$V-I$ colour profile of UGCA~320 after masking out foreground and background sources in the field of view and making corrections for Galactic extinction. The vertical dashed line is the limit of the SALT and MUSE spectroscopic data analysed in this work. The $V-I$ colour profile is very blue and flat at all radii.}
\label{col_prof}
\end{figure}

\section{Results}
\label{results}
\subsection{Stars in UGCA~320}
\subsubsection{Radial colour profile of UGCA~320}
\label{radial_col}
\begin{figure*}
\centering
\includegraphics[width=0.98\textwidth]{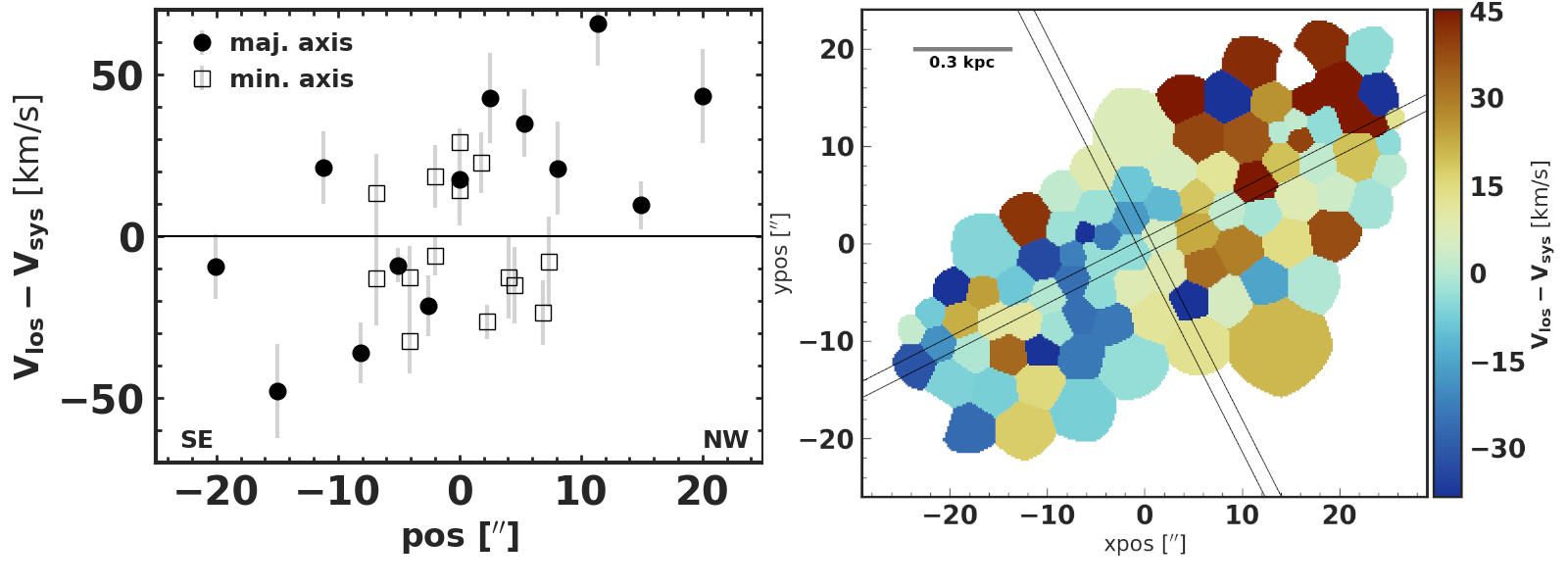}
\caption{{\bf Left panel:} Radial profile of stellar line-of-sight velocities from SALT major (filled circles) and minor (open squares) axes long-slit observations. The horizontal line corresponds to the systemic velocity ($739~\kms$) of UGCA~320. {\bf Right panel:} Stellar line-of-sight velocity map derived from MUSE IFU data analysis. The positions of the SALT long-slits along the major and minor axes are overlaid for reference. North is up and East is left on the map.}
\label{fig:stell_kin_LS_IFU}
\end{figure*}
We obtained $V-I$ radial colour profile from the azimuthally averaged $V~\&~I$-band surface brightness profiles using the 
$\textit{HST}$ imaging data referenced in Section~\ref{sec_HST}. We corrected for Galactic extinction using the maps from \citet{Schlafly2011} and determined the photometric centre with IMFIT \citep{Erwin2015}. The resulting colour profile, shown in Fig.~\ref{col_prof}, extends out to $\unsim90 \arcsec$ ($\unsim2.6$~kpc) well beyond the radial range probed by our spectral data. The radial colour profile is quite blue with an average $V-I$ value of $\unsim0.1$~mag and flat. The blue colour suggests that UGCA~320 is dominated by very young and/or metal-poor stars. The flatness of the colour profile may be due to a lack of radial variation in the stellar population across the optical disc. This lack of a colour gradient also suggests that current star formation in the gas-rich UGCA~320 is not restricted to the galaxy centre. We explore these points later in this work.

\subsubsection{Stellar kinematics in UGCA~320}
\label{stell_kin}
Starting with the centrally extracted global spectra from our LR observations along both the major and minor axes, we measured a mean stellar $\vlos$ of $739 \pm 16~\kms$ consistent within the measurement errors with the systemic velocity of UGCA~320. As mentioned in Section~\ref{data_analysis}, these spectra were extracted from apertures with widths of $7.5 \arcsec$ and $3.5 \arcsec$, respectively. 
We also measured a lowered mean $\vlos$ of $721\pm4~\kms$ and a $\veldis$ of $29\pm5~\kms$ from the HR spectrum (SNR $\unsim16$ per \AA) extracted within a much wider aperture of $37 \arcsec$ ($\unsim 1$~kpc). Our $\veldis$ estimate is consistent with the $\veldis=41$~\kms\ value reported in \citet{Cote2000}, after making corrections for the updated distance of UGCA~320\footnote{\citet{Cote2000} used a distance of $3.5$~Mpc to UGCA~320.}.  

Fig.~\ref{fig:stell_kin_LS_IFU} shows the radial stellar $\vlos$ profile obtained from binned spectra extracted along the major and minor axes of UGCA~320. Due to the exponential fall off of the stellar light, we are only able to obtain stellar kinematics (and stellar population parameters) out to a radius of $\pm 20\arcsec$ ($\unsim0.6$~kpc) along the major axis. Likewise, we show the spatially resolved stellar $\vlos$ map from our MUSE data in Fig.~\ref{fig:stell_kin_LS_IFU}. From both plots, a general sense of how the stellar disc in UGCA~320 rotates can be inferred -- stars are relatively redshifted (blueshifted) in the north-west (south-east) direction, even though the $\vlos$ distribution is irregular and asymmetric. It also appears, especially from the long-slit radial profile, that the kinematics of the stellar disc is heavily disturbed beyond radii $\unsim10\arcsec$ ($\unsim0.3$~kpc) from the galaxy centre. We investigate this further in Section~\ref{kin} using the \texttt{KINEMETRY} method \citep{Krajnovic2006}. 
\\

\begin{figure*}
\centering
\includegraphics[width=0.7\textwidth]{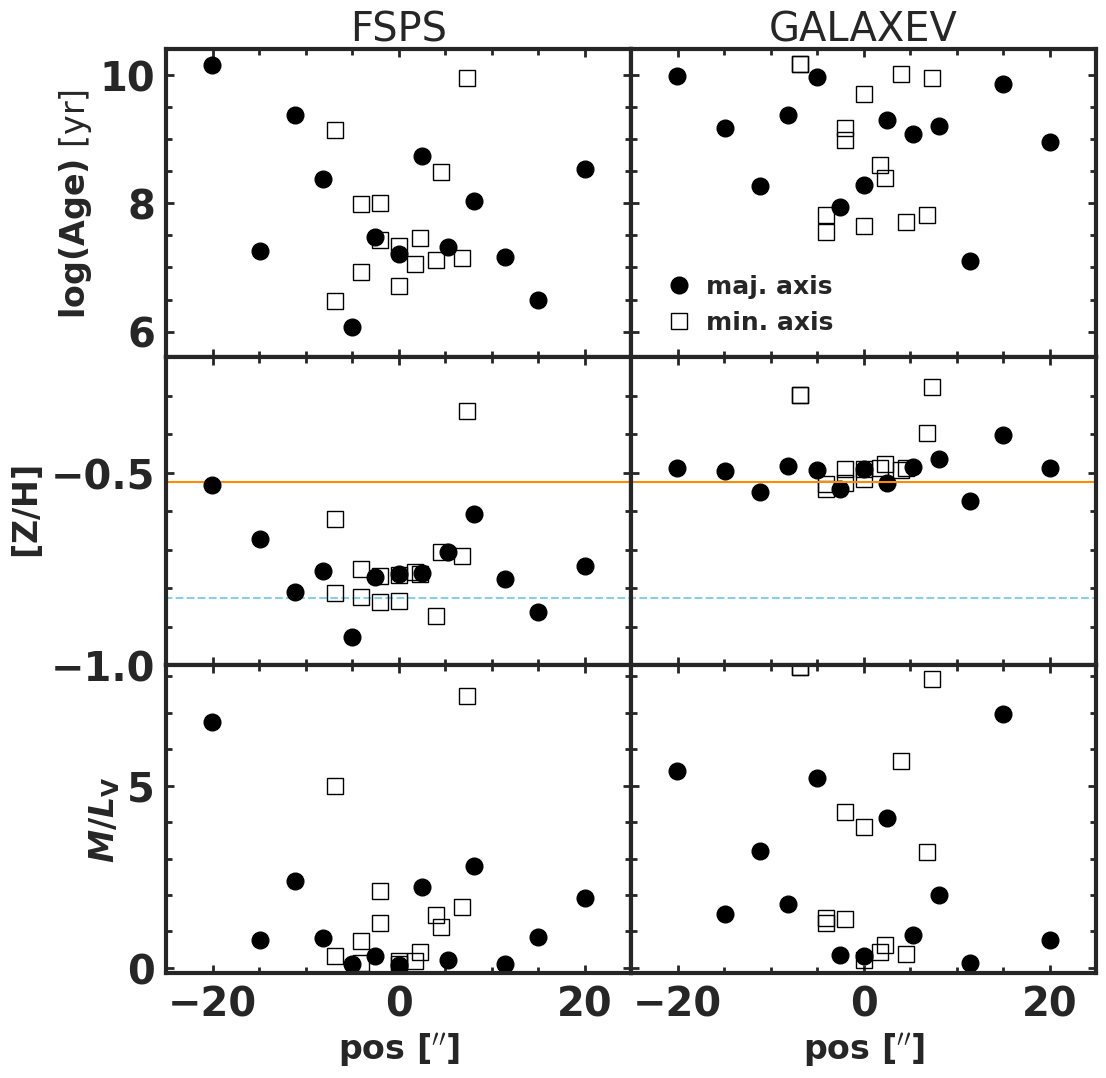}
\caption{Radial profiles of stellar population parameters from \ppxf\ full-spectrum fitting of SALT long-slit data along the major (filled circles) and minor axes (open squares) of UGCA~320. We show the light-weighted logarithmic stellar ages (top panel), the light-weighted stellar metallicities (middle panel), and the $V$-band stellar mass-to-light ratios (bottom panel), obtained using the FSPS synthesis models (left column) and the GALAXEV synthesis models (right column). We also show the $15$ (skyblue/solid line) and the $30$ (darkorange/dashed line) per cent solar metallicity lines in the middle panels to guide the eye.} \label{fig:stell_pop_LS}

\bigskip

\includegraphics[width=0.99\textwidth]{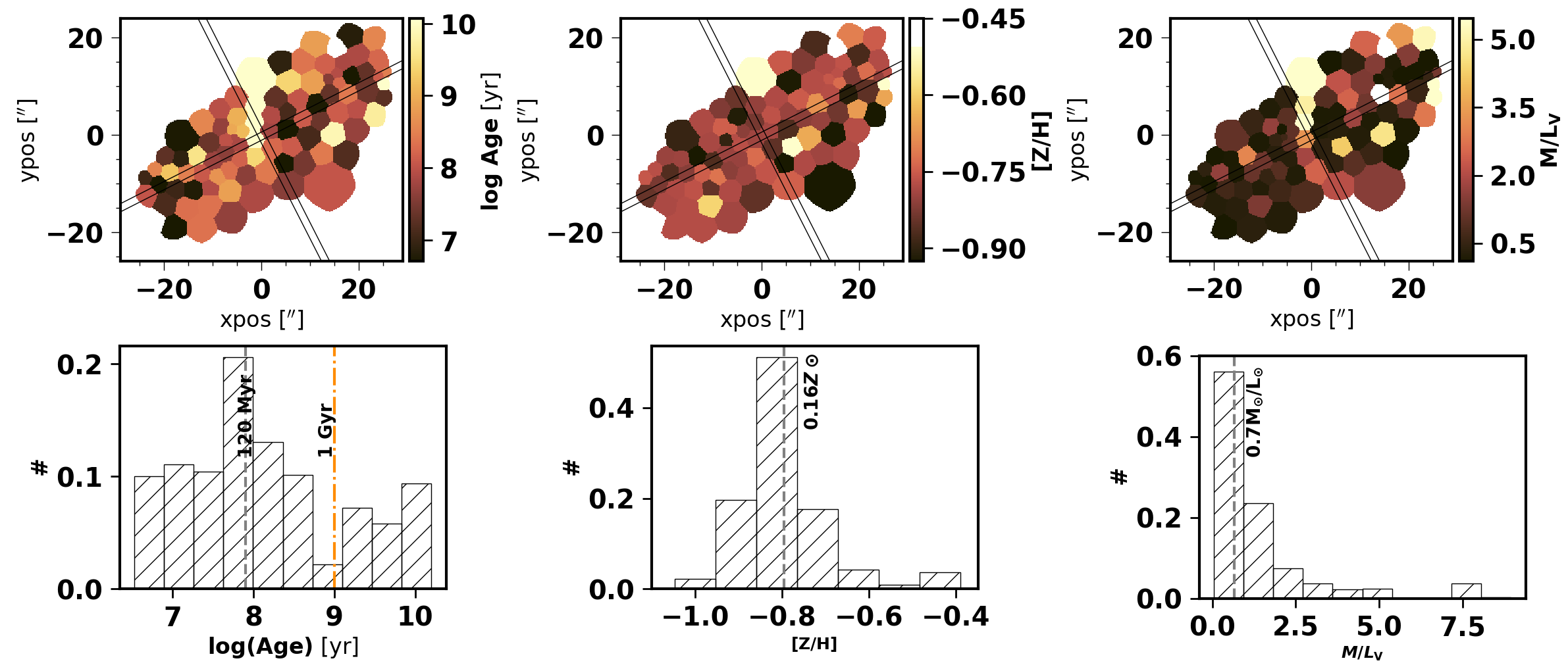}
\caption{Maps of stellar population parameters obtained from \ppxf\ full spectrum fitting of MUSE IFU data with FSPS synthesis model. Top panel, from left to right, shows the light-weighted logarithmic stellar age, the light-weighted stellar metallicities, and the $V$-band stellar mass-to-light ratios. In the bottom panel, we show normalized histograms of the stellar population parameters, marking out the median values with dashed lines. All the maps are overlaid with outlines of the SALT long-slits along the major and minor axes.} \label{fig:stell_pop_IFU}
\end{figure*}

\subsubsection{Stellar population properties of UGCA~320}
\label{stell_pop}
We show radial variations in the stellar age, metallicities, and $M_{*}/L_{\rm v}$ along the major and minor axes in Fig.~\ref{fig:stell_pop_LS}
and their spatial maps in Fig.~\ref{fig:stell_pop_IFU}. The \mtol~values are not obtained directly as fitted parameters from \ppxf\ but rather as estimates derived from the weights of the best fit stellar continuum and the predicted luminosity of each optimal stellar population synthesis (SPS) template in the $V$-band (see eqn. (2) from \citealt{Cappellari2013}). These figures show that the stellar disc of UGCA~320 is indeed dominated by young stars, mostly with light-weighted ages younger than $\unsim1$~Gyr, regardless of the adopted stellar synthesis model. It is, however, important to point out that while the integrated stellar light from the galaxy is dominated by very young stars with a median age of $\unsim120$~Myr, a substantial population of old stars, with light-weighted stellar age $\ge10$~Gyr, is also present in UGCA~320, e.g., the region seen in projection along the minor axis in the northern direction in Fig.~\ref{fig:stell_pop_IFU}. 

Likewise, the stellar disc is metal-poor with a significantly subsolar metallicity (median $\rm [Z/H]\unsim-0.8$~dex, i.e., 
$\unsim15$~per cent solar metallicity from the FSPS model or $-0.5$~dex, i.e., $\unsim30$~per cent solar metallicity with the GALAXEV model) and median $M_{*}/L_{\rm v}$ values of $0.7~\&~0.5~\mtolunit$ with the FSPS and GALAXEV models, respectively. These global $M_{*}/L_{\rm v}$ values agree well with typical values adopted in the literature for dwarf Irregular galaxies of similar stellar mass (e.g., table 3 in \citealt{Kennicutt1994}; fig.~24 in \citealt{Maraston2005}). We remark that similar conclusions about these stellar population parameters can be drawn independently from either the long-slit or IFU spectral data. 

\subsection{Ionized gas in UGCA~320}
\label{gas+}
\begin{figure}
\includegraphics[width=0.48\textwidth]{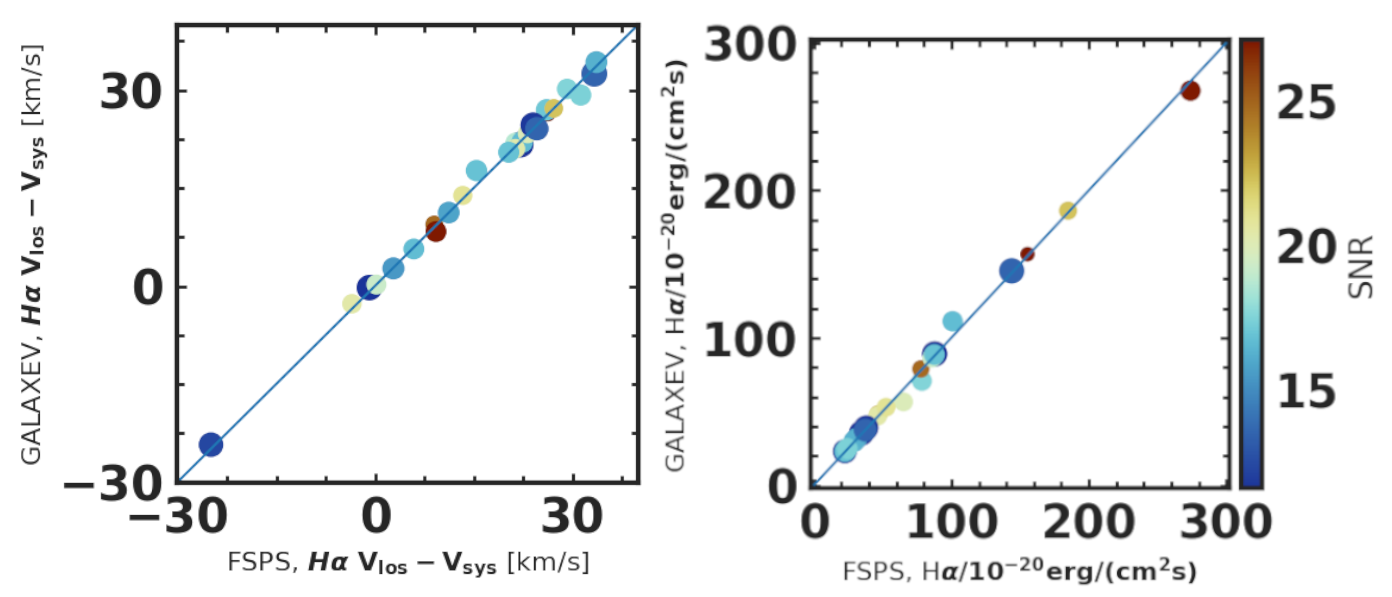}
\caption{Comparison of \Ha\ line-of-sight velocity (left panel) and flux (right panel) measurements from SALT long-slit data using the FSPS and GALAXEV stellar synthesis models. In both plots, the data points are colour-coded by the signal-to-noise ratio in the stellar continuum and their sizes are scaled in proportion to the uncertainties on the flux measurements. We also show the one-to-one line to guide the eye.}
\label{fig:cmp_SPS}
\end{figure}

We begin the characterization of the ionized gas components in UGCA~320 by comparing the measurements made with the two adopted SPS models. Fig.~\ref{fig:cmp_SPS} shows that flux and kinematics measurements made with either of the two models are fully consistent with each other, therefore, in the remaining parts of this paper, we only present ionized gas properties obtained with the FSPS model. 

\subsubsection{Ionized gas flux and their kinematics}
\label{gas_flux_kin}
\begin{figure*}
\centering
\includegraphics[width=0.98\textwidth]{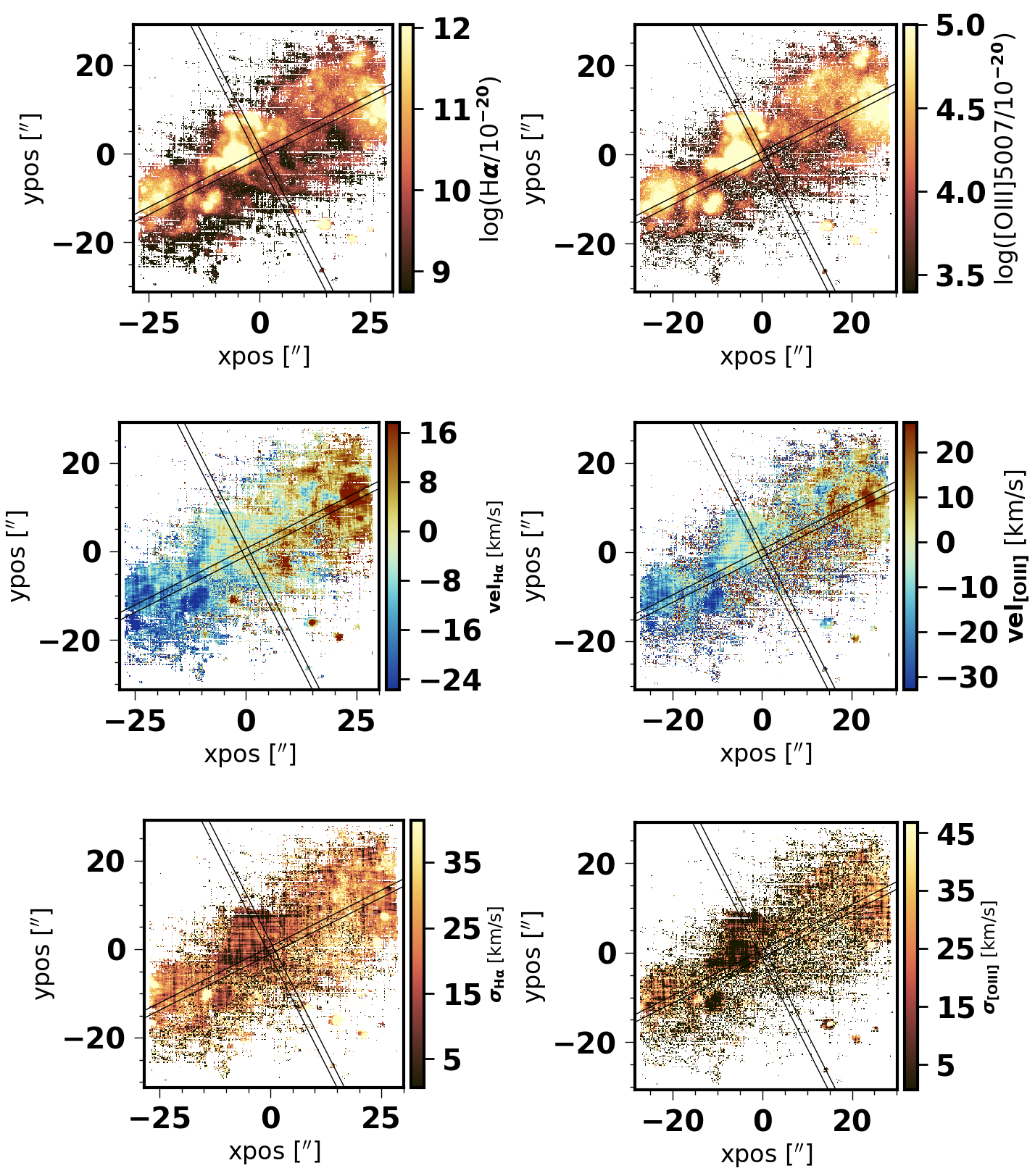}
\caption{Maps of the ionized gas distributions and their kinematics from MUSE IFU data at the native $0.2\arcsec$ spatial resolution. From top to bottom: flux in logarithmic scale and in units of $\rm 10^{-20}~erg~s^{-1}~cm^{-2}~arcsec^{-2}$, line-of-sight velocity fields, and velocity dispersion fields for \Ha\ (left column) \& [OIII] (right column) emission lines, respectively. Ionized gas distribution is clumpy in UGCA~320. Blue and red colours in the middle panel correspond to the approaching and receding sides of the rotating disc. We have overlaid the SALT major and minor axes long-slit outlines as black lines in the maps. North is up and East is left.}
\label{fig:Ionized_IFU}
\end{figure*}

Balmer and forbidden lines are known to be produced by different processes \citep{Burbidge1962}, therefore, a joint examination of the flux distribution and the kinematics of the brightest of these lines from the \HII\ regions in UGCA~320 is appropriate. We show in Fig.~\ref{fig:Ionized_IFU} the ionized gas flux distribution and kinematics of the \Ha\ Balmer line and the [OIII] forbidden line in the FoV of UGCA~320. We remind the reader that we have excluded all spaxels with SNR~$<3$ per \AA\ from these ionized gas maps. From the maps, regions in UGCA~320 with the highest ionized gas intensities are clumpy and are predominantly seen (in projection) above the galaxy's major axis (in the north-east direction) while less intense, more diffuse tails of ionized gas can be seen below the major axis, extending in the south-west direction. This may be due to projection effects as a result of the high inclination of the galaxy ($\unsim80\degr$), such that the north region of the galaxy may be closer to us, hence we observe more ionized gas here along the line of sight compared to the south direction. These individual high flux intensity blobs are typically extended (spanning $\unsim0.1-0.5$~kpc across) and are randomly spread across the entire length of the stellar disc ($\unsim2$~kpc). The major and minor axes slits sample a few of these strong \HII\ regions. The top panel in Fig.~\ref{fig:Ionized_IFU} shows the similarity between the ionized gas distribution as traced by the \Ha\ and the [OIII] fluxes but with a clear domination of the ionized flux budget by the Balmer line emission.

\begin{figure}
\includegraphics[width=0.48\textwidth]{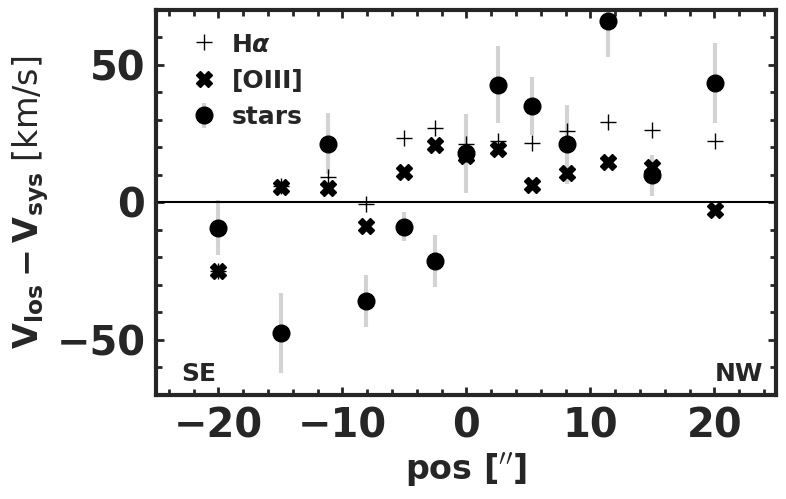}
\caption{Radial profiles of stellar (black filled circles) and ionized gas line-of-sight velocities (\Ha: plus markers; [OIII]: cross markers) from SALT long-slit data along the major axis. Stars and ionized gas in UGCA~320 generally rotate in the same sense -- blueshifted/approaching from the south-east direction and redshifted/receding in the north-west direction.}
\label{fig:rad_kin_prof}
\end{figure}

We show the radial profiles of the ionized gas $\vlos$ along the major axis\footnote{For clarity, we only show the major axis SALT long-slit ionized gas kinematics here.} in Fig.~\ref{fig:rad_kin_prof}. Two things are immediately obvious from the long-slit $\vlos$ data, with additional support from the spatially resolved map of the ionized gas kinematics (shown in Fig.~\ref{fig:Ionized_IFU}): first, the rotation pattern, and secondly, the low amplitude ($\le20~\kms$) of the $\vlos$ on both the receding and approaching sides of the ionized gas disc, especially beyond the central $\unsim10$\arcsec ($\unsim0.3$~kpc) radial region, when compared to the stellar disc ($\le40~\kms$). Furthermore, in the vicinity of the photometric centre of the galaxy\footnote{Here, we assume that the centre of the light profile determined in Section~\ref{gal_centre} corresponds to the photometric centre of the galaxy.}, where $\vlos$ should be consistent with the systemic velocity of the galaxy under equilibrium conditions, the amplitude of the ionized gas $\vlos$ from both the SALT and MUSE data is higher than the $V_{\rm sys}$ by $\unsim10-20~\kms$, suggesting unsettled kinematics. Yet, the $\vlos$ radial profile and the spatially resolved $\vlos$ map of the ionized gas components show coherent motions, with the approaching and the receding kinematics regions clearly discernable (see Fig.~\ref{fig:stell_kin_LS_IFU}). Generally, the ionized gas $\vlos$ increases (decreases) in the north-west (south-east) direction along the major axis within the inner $\unsim10$\arcsec radius, consistent with the stars. Beyond this central region, we see different trends in the ionized gas kinematics on either side of the galaxy centre -- in the south-east direction, the $\vlos$ of the Balmer and forbidden lines show a similar decline while in the north-west direction, we notice a divergence between the kinematics of the Balmer and forbidden lines, albeit, not at a significant level. Together with the disturbed nature of the stellar disc hinted at in Section~\ref{stell_kin}, we explore in more detail the true dynamical state of the stars and ionized gas in UGCA~320 in Section~\ref{kin}.

\begin{figure}
\includegraphics[width=0.48\textwidth]{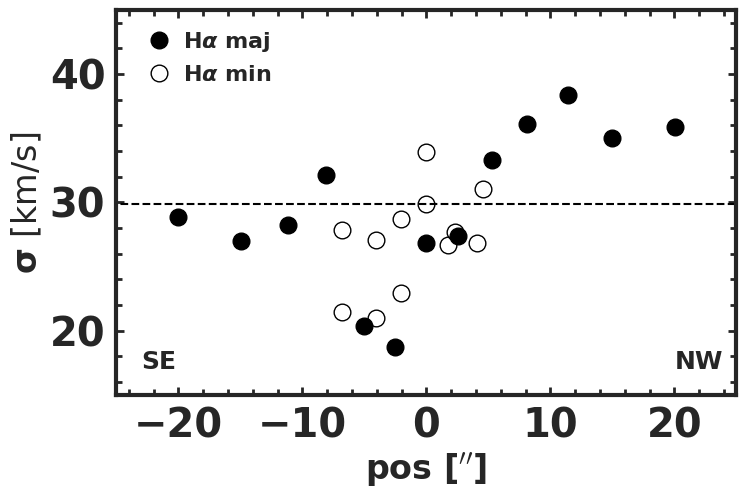}
\caption{Radial profiles of ionized gas velocity dispersion from SALT long-slit data along the major (filled circles) and minor (open circles) axes.
The dashed horizontal line at $\unsim30~\kms$ corresponds to our measured stellar velocity dispersion. Within the limit of our data and measurement uncertainties, stars and ionized gas in UGCA~320 have comparable velocity dispersions.}
\label{fig:vel_disp}
\end{figure}

Examination of the MUSE \Ha\ velocity dispersion field (shown in the bottom panel of Fig.~\ref{fig:Ionized_IFU}) reveals a complex pattern that closely follows the intensity distribution of the ionized gas content in UGCA~320. Gas velocity dispersion is generally suppressed ($\sigma_{\rm H\alpha}\unsim10-20~\kms$) in the high-intensity ionized gas region around the photometric centre of the galaxy (towards the north-east direction). This low $\sigma$ region is bordered on both sides (towards the north-west and south-east directions) by ionized gas clumps with elevated velocity dispersion reaching up to $\sigma_{\rm H\alpha}\sim30-40~\kms$. Ionized gas velocity dispersion is generally elevated in the low flux intensity region below the photometric major axis. We show the \Ha\ velocity dispersion profile from our SALT long-slit data in Fig.~\ref{fig:vel_disp}. Note the dip in $\sigma_{\rm H\alpha}$ to $\unsim20~\kms$ near the galaxy's photometric centre, the rise to $\unsim35~\kms$ at $\unsim10\arcsec$ in the north-west direction, and the rise to $\unsim30~\kms$ in the south-east direction. With the exception of the high flux blob in the north-west direction, where $\sigma_{\rm H\alpha} > \veldis$, ionized gas velocity dispersion in UGCA~320 is comparable to the stellar velocity dispersion.

\subsubsection{Dominant ionization mechanisms}
\label{bpt}
\begin{figure}
\includegraphics[width=0.48\textwidth]{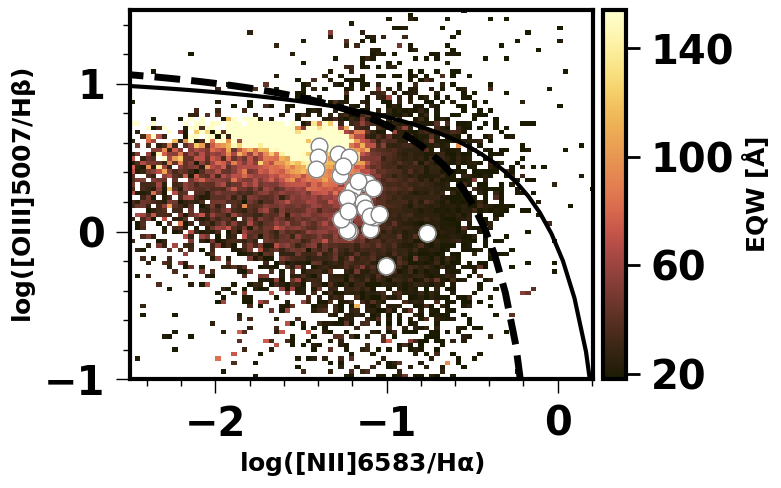}
\caption{BPT diagnostic diagram of all spaxels in UGCA~320. Spaxels from MUSE where \HII\ ionization are powered by recent star formation are found below the solid and the dashed lines from \citet{Kewley2001} and \citet{Kauffmann2003}, respectively. We have colour-coded the spaxels by the \Ha\  equivalent width. The positions of the SALT long-slit spectra extracted along the major and minor axes are shown as white-filled circles and they are all below the diagnostic lines. }
\label{fig:bpt_IFU}
\end{figure}

Having established that \HII\ regions are randomly localized but ubiquitous in UGCA~320, we now address the issue of the mechanism(s) responsible for the gas ionization. We use the BPT (\citealt{Baldwin1981}) diagnostic diagram based on the [OIII]/\Hb\ and the [NII]/\Ha\ ratios with the empirical limits presented in \citet{Kewley2001} and \citet{Kauffmann2003}, respectively, to separate \HII\ regions photo-ionized by UV radiation from recently formed stars from those powered by other processes, e.g., AGN or shocks. Qualitative inferences about the strength of the ionization parameter and the gas-phase metallicity can also be made from the relative position occupied by the \HII\ regions in the BPT parameter space. We show the BPT diagnostic diagram for UGCA~320 in Fig.~\ref{fig:bpt_IFU} and confidently conclude that the source of gas ionization in UGCA~320 is predominantly related to star-forming processes. 

We have also measured the equivalent width of the \Ha\ (hereafter EW) emission line, a well-known proxy for the photo-ionization strength and age of the most recent star forming event \citep{Terlevich2004, Casado2015}. For each spaxel with measured \Ha\ $\vlos$, we fit a single Gaussian profile in rest-frame within a $25$ \AA\ window centred on the \Ha\ feature to obtain the flux intensity which we then divide by the flux contribution from the adjoining continuum region. This, in principle, is a ratio of the ionizing flux from young massive stars to the flux from the underlying older stellar populations. From Fig.~\ref{fig:bpt_IFU}, \HII\ regions in UGCA~320 with the highest EW have an elevated [OIII]/\Hb\ ratio--the maximum possible ratio compatible with ionization powered by recent star-formation at the metallicity of the gas--indicating a high ionization level probably due to presence of young, massive O-type stars. As expected, line flux ratios from our long-slit observations also populate the same region in the BPT digram as the spaxels from the MUSE data, signifying that a similar conclusion can be made independently from either data set. EW measurement from our SALT long-slit data varies from $25-180$~\AA\ along the major and minor axes.

\begin{figure}
\includegraphics[width=0.48\textwidth]{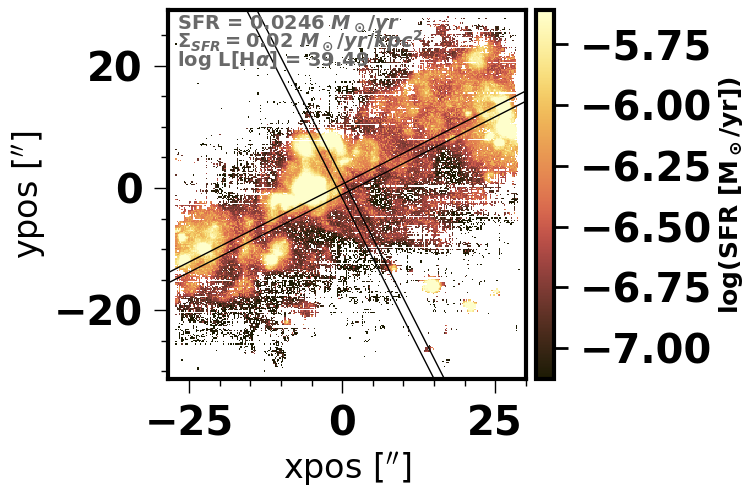}
\caption{Map of star formation rate in UGCA~320 from the star forming regions based on the dust-corrected \Ha\ flux. Overlaid on the map are the outlines of the SALT long-slit along the major and minor axes which miss most of the high star forming regions. Spaxels which fall outside the star-forming region as defined in the BPT diagnostic diagram have been excluded from this plot.}
\label{fig:sfr_IFU}
\end{figure}

\begin{figure*}
\includegraphics[width=0.98\textwidth]{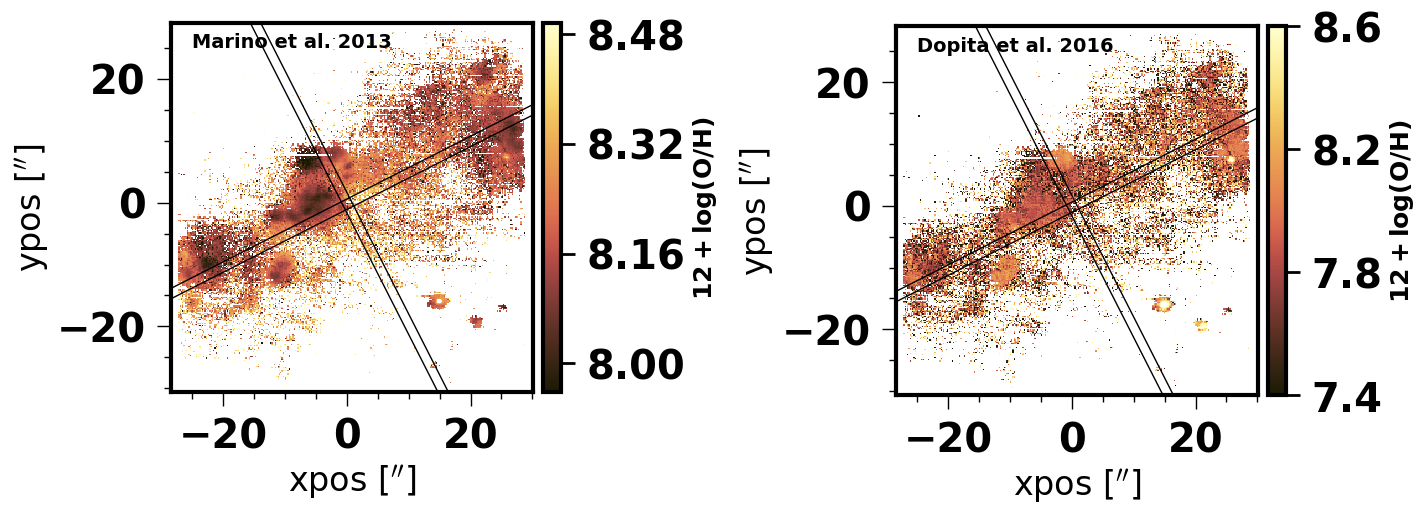}
\caption{Gas-phase metallicity for UGCA~320 using the metallicity abundance calibrators from \citealt{Marino2013} (left panel) and from \citealt{Dopita2016} (right panel). Regardless of the adopted calibrator, UGCA~320 has a gas-phase metallicity that is metal-poor with sub-solar metallicity (see text for more details).}
\label{fig:gas_met}
\end{figure*}

\subsubsection{Star formation rate}
\label{sfr}
The \Ha\ recombination line is a well-known indicator of the current star formation rate in galaxies. We therefore convert the extinction-corrected \Ha\ flux from the MUSE datacube into \Ha\ luminosity, $L(\mathrm{H}\alpha)$, in order to estimate the instantaneous star formation rate (SFR) in UGCA~320 using:
\begin{equation}
    \mathrm{SFR} (M_{\odot} \ \mathrm{yr}^{-1}) = 7.9 \times 10^{-42} L(\mathrm{H} \alpha) \ (\mathrm{erg} \ \mathrm{s}^{-1})
    \label{eq:SFR}
\end{equation} 
from \citet{Kennicutt2009} and show the SFR map in Fig.~\ref{fig:sfr_IFU}. We measure a total $\mathrm{log}~L(\mathrm{H}\alpha)\unsim39.5$, a modest global SFR of $2.5 \times 10^{-2} \ M_{\odot}$ yr$^{-1}$ typical for dwarf Irregulars, a specific SFR (= log(SFR/M$_{*})$) of $-9.5$, and a high SFR density, $\Sigma_{\mathrm{SFR}}$, of $2.0 \times 10^{-2} \ M_{\odot}$ yr$^{-1}$ kpc$^{-2}$ due to the near edge-on inclination of UGCA~320, obtained by dividing the SFR by the area (in kpc$^2$) of the contributing spaxels. We remind the reader that the faint, extended \HII\ region in the north-west direction (see Section~\ref{sec_HST}) is not included in the results presented here, but we estimate, using the flux values presented in the table~3 from \citet{Cote2009}, that this unavailable region contributes at most $10$~per cent of the total \Ha\ flux budget of the galaxy. This addition only alters our $\mathrm{log}~L(\mathrm{H}\alpha)$ estimate by $0.05$~dex. 

While a direct comparison of our spectroscopically measured global $\mathrm{log}~L(\mathrm{H}\alpha)$ and SFR with estimates from the literature (e.g., table~3 from the compilation in \citealt{Kennicutt2008}; table~2 from \citealt{Cote2009};  table~A1 from \citealt{Karachentsev2021}) is not straightforward due to systematics imposed by differences in SFR tracers, data type, extent and depth, adopted methods (compare eqn. 1 in \citealt{Cote2009} with our eqn~\ref{eq:SFR}), etc., it is safe to state that the global star-forming activity in UGCA~320 is low regardless of data or method, and typical of estimates derived for other nearby dwarf irregular galaxies in the literature. This SFR is, however, adequate to produce the total stellar mass of the galaxy within a Hubble time. We estimate a gas depletion timescale\footnote{The star formation efficiency is the inverse of this term.} of $\unsim49.5$~Gyr for UGCA~320, defined as the number of years it will take the galaxy to exhaust its entire gas reservoir, assuming star formation continues at the current SFR (total gas mass/SFR)\footnote{We estimate the total gas mass as $1.4 \times M\mathrm{(HI)}$ to account for He and use $M\mathrm{(HI)}$ from table 1 in \citet{deBlok2024}.} and that all of the cold gas can migrate towards the disc.

\subsubsection{Gas-phase metallicity}
\label{gas_met}
Results from our stellar population analysis in Section~\ref{stell_pop} in addition to the very blue colour from Section~\ref{radial_col} already show that the stellar content of UGCA~320 is metal-poor. Being a gas-rich galaxy, we now examine the metallicity of the ionized gas content. There are many gas-phase metallicity (Z$_{g}$) calibrators in the literature \citep[e.g.,][]{Kewley2008} that are based on the relative abundance of oxygen (e.g., \citealt{Marino2013}, hereafter M13; \citealt{Dopita2016}, hereafter D16). Here, we present oxygen abundance estimates from calibrators that are applicable to the wavelength range of our spectral data based on our dereddened \Ha, \Hb, [OIII], [NII] and [SII] flux measurements. The calibrators we use are:
\begin{equation}
\label{M13}
\mathrm{12 + log (O/H)} = 8.533-0.214 \times \mathrm{log} \bigg(\frac{\mathrm{[OIII]}}{\mathrm{H}\beta} \times \frac{\mathrm{H}\alpha}{\mathrm{[NII]}}\bigg)  \tag{2, M13}\\
\end{equation}  
and 
\begin{equation}
\label{D16}
\mathrm{12 + log (O/H)} = 8.77 + \mathrm{log} \bigg(\frac{\mathrm{[NII]}}{\mathrm{[SII]}} \bigg) + 0.264 \times \bigg(\frac{\mathrm{[NII]}}{\mathrm{H}\alpha} \bigg)  \tag{3, D16}.\\
\end{equation}
We show the Z$_{g}$ maps in Fig.~\ref{fig:gas_met} and note that Z$_{g}$ estimated with the M13 and D16 calibrators range from $ 8.0 < \mathrm{12 + log (O/H)} < 8.4$ and $ 7.4 < \mathrm{12 + log (O/H)} < 8.6$, respectively. Adopting a Solar metallicity with $\mathrm{12 + log (O/H)} \unsim 8.7$ \citep{Asplund2009,Bergemann2021,Pietrow2023} implies that the Z$_{g}$ of UGCA~320 is firmly subsolar with median oxygen abundances $\mathrm{12 + log (O/H)} \unsim 8.2$ (M13) and $\unsim7.9$ (D16) corresponding to $\unsim 0.3$ and $\unsim 0.15$ Solar metallicity, respectively. \citet{Lee2003} in a study of southern dwarf galaxies reported a gas-phase metallicity of $\mathrm{12 + log (O/H)} \unsim 8.06$ for UGCA~320, which compares well with the median Z$_{g}$ values we find in this work. Our median Z$_{g}$ are comparable with the stellar metallicities (Z$_{*}$) obtained in Section~\ref{stell_pop}, highlighting the natural link between the stellar and ionized gas components of UGCA~320. 

A careful look at the Z$_{g}$ maps in Fig.~\ref{fig:gas_met} reveals significant structure in Z$_{g}$ within individual HII regions from the M13 calibrator but not seen in the version made with the D16 calibrator. This artefact of the M13 calibrator has already been noted in the literature, e.g., \citet{Kruhler2017}, where it is claimed to be due to the calibrator's sensitivity to local changes in the ionization parameter. Overall, there is no gradient in the global gas-phase metallicity maps of UGCA~320 from both calibrators.

\subsection{Kinemetry}
\label{kin}
As mentioned in Section~\ref{sec_HST}, the morphological perturbation of UGCA~320 is already evident from deep and wide-field imaging which reveals a very faint, narrow and extended stellar structure in the north-west direction (see Fig~\ref{fig:colImg}). Here, we present a more detailed analysis of the 2D stellar and ionized gas velocity fields in order to gain more insight into the dynamical state of UGCA~320. We use the \texttt{KINEMETRY} method, developed by \citet{Krajnovic2006}\footnote{This method builds on earlier works introduced by \citet{Franx1994,Schoenmakers1997}}, which describes our velocity fields as a series of concentric ellipses which satisfies a simple cosine law, each of which is centred on the galaxy's kinematic centre (which may or not be the same as the photometric centre) and further parametrized by a kinematic position angle (PA$_{kin}$)\footnote{This is defined as the position angle of the maximum velocity, i.e., receding velocity, within each concentric ellipse, and it is measured from East towards the North.} and flattening ($q = b/a$). 
\begin{figure*}
\includegraphics[width=0.80\textwidth]{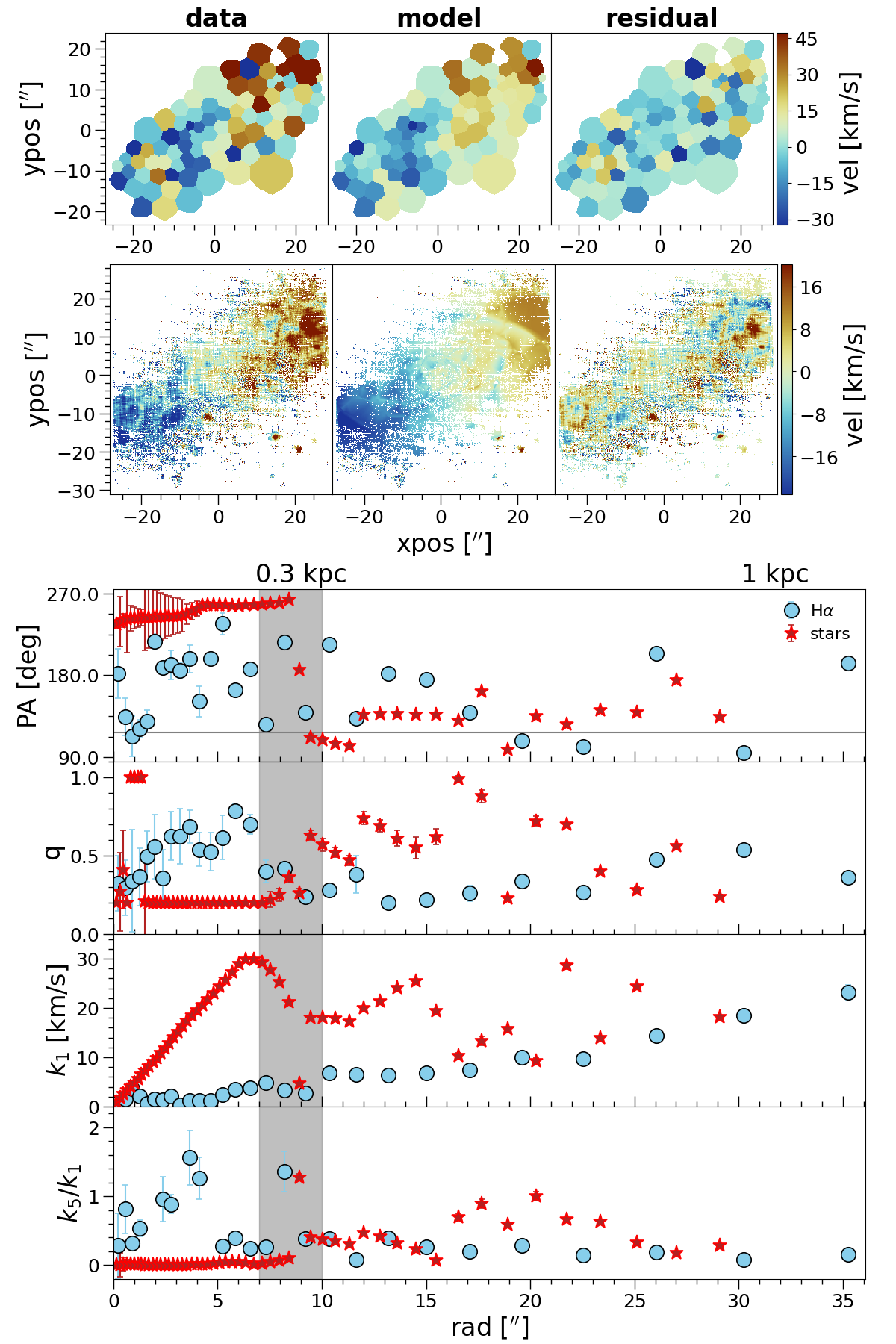}
\caption{\textbf{Upper panel:} Voronoi-binned map of stellar line-of-sight velocities (top left), map of \Ha\ line-of-sight velocities (bottom left) and their respective models obtained from \texttt{KINEMETRY} fitting (middle) and residual (data $\minus$ model) maps (right) for UGCA~320. \textbf{Lower panel:} Radial profiles of the kinematic properties from \texttt{KINEMETRY}. Stars are shown as red-filled stars and \Ha\ as blue-filled circles. From top to bottom, we show radial profiles of the kinematic position angle (PA), the flattening ($q$), the $k_{1}$ and the $k_{5}/k_{1}$ parameters, respectively. The photometric PA is shown in the top plot with the gray line for comparison. Note the sharp transition in the properties of the stellar kinematics at $\unsim 10\arcsec$ ($\unsim0.3$~kpc) marked by the vertical grey band.}
\label{fig:kin_plots}
\end{figure*}
For each ellipse, the kinematic moment in the plane of the sky, $K(\rm{r},\psi)$, can be represented as a Fourier series with a finite (typically a few) number of harmonic terms of the form: 
\begin{equation}
\mathrm{K(r, \psi) = A_{0}(r)} + \sum _{n=1}^N A_{n}\mathrm{(r) \hspace{1mm}sin(n\psi) + B_{n}(r)\hspace{1mm} cos(n\psi)} 
\end{equation} 
or in a more compact form as 
\begin{equation}
\mathrm{K(r, \psi) = A_{0}(r)} + \sum _{n=1}^N k_{n}\mathrm{(r) cos[n(\psi-\phi_{n}(r))]}
\label{kin_eqn}
\end{equation} 
where $r$ is the semi-major axis of the ellipse, $\psi$ is the azimuthal angle, measured from the projected major axis, in the plane of the galaxy.
The $k_{n}$ and $\phi_{n}$ parameters are the amplitudes and phase coefficients of each harmonic term and are defined as:
$k_{n} = \sqrt{A_{n}^{2} + B_{n}^{2}}$ and $\phi_{n} = \mathrm{arctan} \left(\frac{A_{n}}{B_{n}}\right)$. 
In the case of our velocity fields, which are odd moments of the LOS velocity distributions, $A_{0}$, the zeroth-order term, is the systemic velocity of each ellipse, which is $\unsim0~\kms$ by construction. The important parameters of interest obtained from the \texttt{KINEMETRY} modelling are the best-fit PA$_{\rm kin}$, $q$, the $k_{1}$ and $k_{5}/k_{1}$ parameters from each ellipse\footnote{$k_{3}$, which is not presented in our analysis, is mostly zero as expected for a well determined fit (see sec. 2.2 in \cite{Schoenmakers1997}).}. The first two of these parameters are straightforward and their radial profiles are key in identifying substructures in the velocity field. The $k_{1}$ parameter is simply the best-fit rotation amplitude for each ellipse, and as we show below, it is very crucial in properly discerning trends in the often irregular stellar velocity fields of dwarf Irregular galaxies. The $k_{5}$ parameter captures and quantifies significant deviation from simple rotation in the velocity field where the expectation is that the velocity field peaks at the major axis and goes to zero at the minor axis.

We show the results from the \texttt{KINEMETRY} analysis of the stellar and \Ha\ velocity fields in Fig.~\ref{fig:kin_plots}. We show the original data, the reconstructed velocity models, and the residual maps (i.e., data -- model) with typical rms of $\unsim14~\kms$ and $\unsim9~\kms$ in the stellar and \Ha\ maps, respectively. We also show the radial profiles of the parameters introduced earlier. Despite the coarseness of the stellar velocity field (due to spatial binning) compared to the \Ha\ velocity field where we retain the original spatial resolution of MUSE, we obtained stable fits to all the \texttt{KINEMETRY} parameters. As shown in Fig.~\ref{fig:kin_plots}, we find a sharp transition in the stellar \texttt{KINEMETRY} parameters at $r \unsim 10\arcsec$. In the inner region, i.e., $r \le 10\arcsec (\unsim 0.3$~kpc), the PA$_{\rm kin}$ is offset at $\unsim 250 \degr$ compared to the photometric PA, but beyond this transition radius, the PA$_{\rm kin}$ drops sharply to values that are generally consistent with the photometric PA. We observe similar radial trends for the other parameters: $q$ varies mostly from $0.2-0.4$ in the inner region and beyond, rises up to $\unsim1$; $k_{1}$ increases steadily with radius up to a peak value of $\unsim 30~\kms$ near the edge of the inner region, and beyond, drops and varies from $\unsim10-20~\kms$ out to large radii; $k_{5}/k_{1}$ is stable mostly at $\unsim0$ within $r \le 10\arcsec$ signifying the presence of a regular, disc-like rotation, but beyond rises in a noisy, unstable way. The shape of the stellar disc's $k_{1}$ radial profile is reminiscent of the rotation curves of galaxies undergoing tidal interactions e.g., \citet{Barton1999}. These authors claim from their $N$-body simulations that during the early stages of the tidal interaction between a galaxy pair, the stellar disc segregates into two distinct kinematic components -- a central bar-like structure and a distorted outer disc.    

Our \texttt{KINEMETRY} analysis of the ionized gas velocity field does not show any dramatic transition in \texttt{KINEMETRY} parameters around $r \unsim 10\arcsec$, suggesting a decoupling between the between the kinematics of the stars and the ionized gas. Generally, trends in the radial profiles of the fitted \texttt{KINEMETRY} parameters of the ionized gas show more fluctuations compared to what we obtained from the stellar velocity field, especially in the most central region. PA$_{\rm kin}$ fluctuates rapidly but is always less than the photometric and the stellar kinematic PA within $r \le 8\arcsec$, and it drops off to lower values at larger radii. $q$ increases steadily from $\unsim0.2$ to $\unsim0.8$ in the inner region and drops off to lower values at larger radii. The \Ha\ rotation amplitude, $k_{1}$, is very low, i.e.,$\le 5~\kms$, compared to the stellar amplitude in the inner region and only reaches $\unsim 10~\kms$ at $r\unsim20\arcsec$, in agreement with our earlier inference from the long-slit kinematics profile. We confirm the accuracy of this measurement by visually comparing with the gas-phase velocity radial profile (see fig. 2 for DDO~161 as published in \citet{Cote2000}) where comparable amplitudes are reported in the innermost 0.5~kpc region. We, however, defer a future examination of the origin and implications of this probable decoupling to a future study when deep, high resolution spectral data becomes available. Unlike the radial profile for the stellar rotation amplitude, $k_{1}$ for \Ha\ continues to rise even out to our most distant radial position. Lastly, we note that the $k_{5}/k_{1}$ rises to values $>1$ in the inner region, opposite to the trend observed for the stellar disc while settling to $\unsim0$ at larger radii.

\section{Discussion and Conclusion}
\label{discuss}
\begin{figure*}
\includegraphics[width=0.98\textwidth]{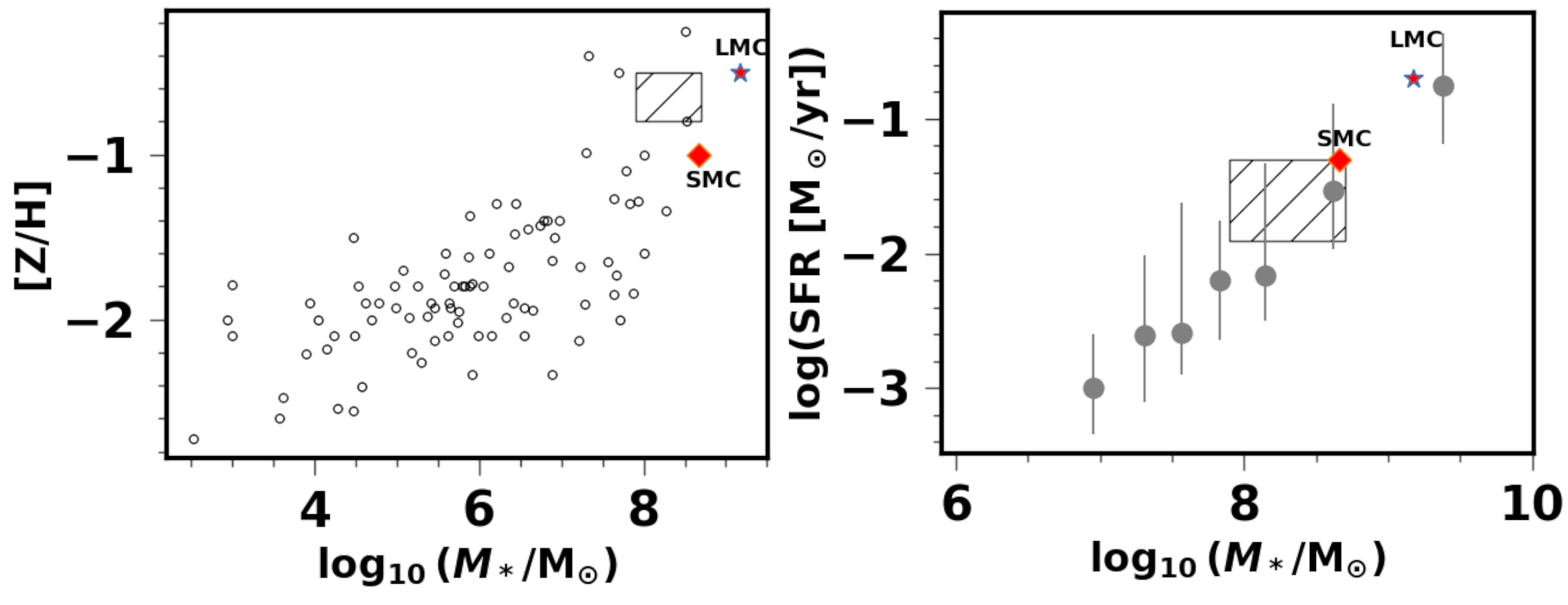}
\caption{\textbf{Left panel:} Comparison of the metallicity (both stellar and gas-phase metallicities) of UGCA~320 with dwarf galaxies (open circles) in the Local Group using the compilation from \citet{McCon2012}, and \textbf{Right panel:} the star formation rate (SFR) with the SF sequence of satellite systems around MW-like galaxies from the SAGA Survey \citep{Geha2024}. UGCA~320 is represented with shaded boxes in both plots to reflect the span of measured metallicities, the range of stellar masses reported in the literature, and the possible range of SFR assuming a factor of two enhancement due to tidal interaction with its neighbour \citep{Stierwalt2015}. In both plots, we explicitly show how UGCA~320 compares with the LMC (filled star) and the SMC (filled diamond).}
\label{fig:cmp_LG}
\end{figure*}
Our analysis in Section~\ref{results} shows that the optical $V-I$ colour of UGCA~320 is very blue. This blue colour may be due to the domination of the stellar population by young and/or metal-poor stars. But as argued for the Magellanic Irregulars \citep[e.g.,][]{Tollerud2011}, the colour may also bear the imprints of recent star formation triggered by tidal forces during the group assembly or infall towards the larger group dominated by NGC~5068. In order to shed more light on the origin of this blue colour, we have obtained via full spectrum fitting (Section~\ref{stell_pop}) and the gas-phase metallicity analysis (Section~\ref{gas_met}), the age and metallicity of the stars (see the histograms in Fig.~\ref{fig:stell_pop_IFU}) as well as the gas-phase metallicity of UGCA~320 (see Fig.~\ref{gas_met}). UGCA~320 is clearly dominated by young stars with ages $< 1$~Gyr even though it hosts a significant subpopulation of old stars (age $> 10$~Gyr). The measured metallicities in this work, from the stars (Z$_{*}$) and the ionized gas-phase ISM (Z$_{g}$), are subsolar, respectively, ranging from $15-30$~per cent of the solar value. For comparison, the LMC and the SMC pair have subsolar gas-phase metallicities that are $\unsim47$ and $\unsim20$ per cent of the solar value, respectively \citep{Dom2022}.
 
Furthermore, there is no substantial difference between our measured Z$_{*}$ and Z$_{g}$ for UGCA~320. This is at odds with observational results from \citet{Gallazzi2005,Fraser2022}, where Z$_{g} > $ Z$_{*}$, with the difference becoming larger in low-mass galaxies. We remind the reader that Z$_{*}$ is an average of the metal-enrichment history of the galaxy's ISM that is incorporated into the stars when they are formed while Z$_{g}$ is the instantaneous value of the metal-enrichment level retained in the ISM, thus, they trace different timescales. We also note that the integrated starlight from UGCA~320 is dominated by young stars (see the histograms in Fig.~\ref{fig:stell_pop_IFU}), hence the close match between both metallicity parameters. In fact, the measured metallicities are consistent with the stellar mass--metallicity relation for dwarf galaxies in the Local Group, using the mass--metallicity compilation from \citet{McCon2012} and assuming that $\rm [Z/H] \unsim [Fe/H]$ (see Fig.~\ref{fig:cmp_LG}). In this regard, therefore, UGCA~320 is not an unusual galaxy.

Initial results from the deep HI MHONGOOSE survey of UGCA~320 reveal an extended, asymmetric HI field but show no direct gaseous bridge connecting it to any of its neighbours. This is not an unusual observation in interacting galaxies that are just approaching or passing the apocentre for the first time \citep{Moreno2015}. As mentioned in Section~\ref{sec_HST}, the extended and asymmetric starlight component in the north-west direction along the major axis of UGCA~320 suggests a tidal interaction. Similar asymmetric features have been observed in the outer discs of nearby dwarf Irregular galaxies \citep{Hunter2011} and they are believed to be artefacts of tidal interactions \citep{Cardiel2024}. Interestingly, this asymmetric stellar feature is embedded (co-spatial) in the distorted region of the HI field, suggesting that they may share the same origin. Unfortunately, we are unable to spectroscopically probe this region in this work due to the limited FoV from MUSE and SALT and the faintness of the stars and \HII\ regions in the outer discs. The disturbed optical morphology of UGCA~320 suggests that its evolution has been significantly impacted by its interaction with the environment.

The nearest neighbour to UGCA~320 is UGCA~319, seen in the north-west direction in the plane of the sky, at a physical separation of $\unsim33$~kpc and a velocity separation of $\unsim12~\kms$. Gas-rich, low-mass galaxy pairs this close are expected to experience a factor of $\unsim2.5$ boost in their SFR \citep{Stierwalt2015}. This is because close interactions can lead to an increase in gas densities through compressions which then trigger star formation. Clearly, our SFR analysis in this work (see Section~\ref{sfr}), together with previous estimates from the literature \citep{Meurer2006, Cote2009, Leroy2019} show that UGCA~320 is not a starbursting galaxy. Neither is UGCA~319 starbursting\footnote{Star formation rate for LEDA~886203 is unavailable.} \citep{Karachentsev2017}. We have measured a SFR of $2.5 \times 10^{-2} \ M_{\odot}$ yr$^{-1}$ which is lower when compared to that of the LMC ($\unsim0.2 \ M_{\odot}$ yr$^{-1}$ - \citealt{Whitney2008}) and the SMC ($\unsim0.05 \ M_{\odot}$ yr$^{-1}$ - \citealt{Wilke2004}) pair which are not starbursting but are believed to be in the early stage of their tidal interaction \citep{Besla2012}. UGCA~320 has a SFR that is typical for dwarf Irregular galaxies with its gas fraction and is consistent with the $\mathrm{M_{*} - SFR}$ relation (see Fig.~\ref{fig:cmp_LG}) where we compare with data from the SAGA Survey of satellite galaxies around MW-type galaxies \citep{Geha2024}. As shown in Fig.~\ref{fig:cmp_LG}, even if the present-day SFR of UGCA~320 has been enhanced via its interaction with its neighbours, what is evident based on the high gas fraction (M$_{\rm HI}$/M$_{*}\unsim2-10$ from Table~\ref{tab:globprop}), low stellar and gas-phase metallicities, and modest SFR of UGCA~320 is that, chemically, it is a slowly evolving galaxy. 

The kinematics of UGCA~320 is peculiar and complex. Rotation is observed in both the stellar and ionized gas components, with both rotating in the same sense and having comparable velocity dispersion (at least along the major axis). However, within the innermost $\unsim10\arcsec$ ($\unsim0.3$~kpc) region, the rotation amplitude of the stars rises steeply to $\unsim30~\kms$, more than that of the ionized gas component which is $\le10~\kms$. In fact, at any radii across the entire disc region probed in this work, the rotation amplitude of the ionized gas is always lower than that of the stars (consistent with the rotation profiles presented in fig. 2 of \citealt{Cote2000}). Stars and ionized gas by their nature respond to gravitational perturbations differently and on varying timescales, the former being collisionless and on longer timescales and the latter collisional and on shorter timescales \citep{Eggen1962}. In this sense, stars, unlike ionized gas, are better at retaining memories of significant interactions. 

\citet{Bagge2024} recently showed that signatures of tidal interactions are more pronounced and last longer in the stars (rather than the ionized gas component) of low-mass galaxies which have young mean stellar ages. This is consistent with the results presented in Section~\ref{kin}. It has been argued in the literature (see the discussion in \citealt{Ebrova2021} for a summary) that kinematically decoupled cores, such as we have found in UGCA~320, could be formed from gas accreted on retrograde orbits, gas-rich minor merger or tidal perturbations from flyby of neighboring galaxies. The first two processes are expected to also leave 
imprints in the chemical composition of the remnant galaxy. However, our results in this work show that the kinematically distinct inner and outer stellar disc components do not correspond to unique stellar population components (see Figs.~\ref{fig:stell_pop_LS} and \ref{fig:stell_pop_IFU}), thus lending more credence to the tidal interaction origin of UGCA~320.

A follow-up IFU spectroscopic study of UGCA~319 which is key in developing a complete evolutionary picture of this dwarf galaxy group is already planned. Such a study will help confirm if the disturbances in the morphology and kinematics of UGCA~320 are mutual and if they are due to their tidal interaction. It is also important to spectroscopically study the extended asymmetric stellar feature in the north-west direction (see Fig.\ref{fig:colImg}) since it could be the remnant of a recently cannibalized companion. 


\section*{Acknowledgements}
We thank the reviewer for their thorough and constructive comments, and suggestions that improved the paper.
SIL is supported in part by the National Research Foundation (NRF) of South Africa (CPRR240414214079). Any opinion, finding, and conclusion or recommendation ex- pressed in this material is that of the author(s), and the NRF does not accept any liability in this regard. 
NZ is supported through the South African Research Chairs Initiative of the Department of Science and Technology and National Research Foundation.
All long-slit spectra observations reported in this paper were obtained with the South African Large Telescope (SALT) under programme numbers 2020-2-SCI-029 and 2021-1-MLT-002 (PI: Mogotse). Any opinion, finding, and conclusion or recommendation expressed in this material is that of the author(s), and the NRF does not accept any liability in this regard. 
This research has made use of the services of the ESO Science Archive Facility. 
This research is based on observations collected at the European Southern Observatory
under ESO programme ID 105.20GY. 
This research used Astropy,\footnote{http://www.astropy.org} a community-developed core Python package for Astronomy \citep{Astropy2013, Astropy2018}.
This research has made use of the NASA/IPAC Extragalactic Database (NED), 
which is funded by the National Aeronautics and Space Administration and operated by the California Institute of Technology.
This research has made use of \texttt{TOPCAT},\footnote{http://www.starlink.ac.uk/topcat/} \citep{Taylor2005}.
The Pan-STARRS1 Surveys (PS1) and the PS1 public science archive have been made possible through contributions by the Institute for Astronomy, the University of Hawaii, the Pan-STARRS Project Office, the Max-Planck Society and its participating institutes, the Max Planck Institute for Astronomy, Heidelberg and the Max Planck Institute for Extraterrestrial Physics, Garching, The Johns Hopkins University, Durham University, the University of Edinburgh, the Queen's University Belfast, the Harvard-Smithsonian Center for Astrophysics, the Las Cumbres Observatory Global Telescope Network Incorporated, the National Central University of Taiwan, the Space Telescope Science Institute, the National Aeronautics and Space Administration under Grant No. NNX08AR22G issued through the Planetary Science Division of the NASA Science Mission Directorate, the National Science Foundation Grant No. AST-1238877, the University of Maryland, Eotvos Lorand University (ELTE), the Los Alamos National Laboratory, and the Gordon and Betty Moore Foundation.
\section*{Data Availability}
The reduced MUSE data that support the findings of this study are
available in the ESO Science Archive at http://archive.eso.org/cms.
html and can be accessed with programme ID 105.20GY.



\bibliographystyle{mnras}
\bibliography{UGCA320} 



\bsp	
\label{lastpage}
\end{document}